\documentclass[useAMS,usenatbib,onecolumn]{mnras}

\usepackage{dcolumn}
\usepackage{physics}
\usepackage{adjustbox}
\usepackage{graphicx}
\usepackage{verbatim}
\usepackage{tikz}
\usepackage{caption}
\usepackage[version=4]{mhchem}
\usepackage{xcolor}
\usepackage{url}
\usepackage{color}

\usepackage{epsf}
\usepackage{amsmath}
\usepackage{lscape}	
\usepackage{bm}

\usepackage{siunitx}

\usepackage{booktabs}
\usepackage{multirow}

\usepackage{dcolumn}
\newcolumntype{d}[1]{D{.}{.}{#1}}

\newcommand{\cm}{cm$^{-1}$}

\newcommand{\ai}{\textit{ab initio}}

\newcommand{\trove}{\textsc{trove}}
\newcommand{\TROVE}{{\sc TROVE}}
\newcommand{\Marvel}{{\sc Marvel}}

\newcommand{\name}{CoYuTe-15}

\graphicspath{{img/}}

\newcommand{\Dh}[1]{${\mathcal D}_{#1{\rm h}}$}

\newcommand{\NH}[2]{$^{#2}$NH$_{#1}$}

\newcommand{\hl}[1]{#1}

\title[ExoMol line lists -- LX. $^{15}$NH$_3$]{ExoMol line lists -- LX.  Molecular line list for the ammonia isotopologue $^{15}$NH$_3$}

\date{\today}

\author[Yurchenko et al.]{Sergei N. Yurchenko\thanks{The corresponding author: s.yurchenko@ucl.ac.uk},
Charles A. Bowesman,
Ryan P. Brady,
Elizabeth R. Guest,
Kyriaki Kefala,
\newauthor{Georgi B. Mitev,
Alec Owens,
Armando N. Perri,
Marco Pezzella,
Oleksiy Smola,}
\newauthor{Andrei Sokolov,
Jingxin Zhang, Jonathan Tennyson\thanks{The corresponding author: j.tennyson@ucl.ac.uk}}\\
Department of Physics and Astronomy, University College London, Gower Street, WC1E 6BT London, United Kingdom}

\date{Accepted XXXX. Received XXXX; in original form XXXX}

\pagerange{\pageref{firstpage}--\pageref{lastpage}} \pubyear{2024}

\begin{document}

\label{firstpage}

\maketitle

\begin{abstract}

A theoretical line list for $^{15}$NH$_3$  CoYuTe-15 is presented based on the  empirical potential energy  and \textit{ab initio} dipole moments surfaces developed and  used for the production of the ExoMol line list CoYuTe  for $^{14}$NH$_3$. The ro-vibrational energy levels and wavefunctions are computed using the variational program TROVE. The line list ranges up to 10~000~cm$^{-1}$ ($\lambda \geq 1$~$\mu$m) and contains  929~795~249 transitions between 1~269~961 states with $J\le 30$. The line list should be applicable for temperatures up to $\sim$1000~K. To improve the accuracy of the line positions, a set of experimentally-derived energy levels of $^{15}$NH$_3$ is produced using the \Marvel\ procedure. To this end,  37 experimental sources of the line positions of $^{15}$NH$_3$ available in the literature are collected, combined and systematised to produce a self-consistent spectroscopic network of \hl{21~095} $^{15}$NH$_3$ transitions covering \hl{40} vibrational bands ranging up to \hl{6818} cm$^{-1}$ and resulting in \hl{2777} energy term values. These \Marvel\ energies are then used to replace the theoretical values in the CoYuTe-15 line list and  also complemented by pseudo-\Marvel\ energies obtained by an isotopologue  extrapolation using the previously reported \Marvel\ energies of the $^{14}$NH$_3$ parent isotopologue of ammonia. A list of 53~856 high resolution transitions between \Marvel\ states and theoretical intensities is provided in the HITRAN format. Comparison with the recent experimental spectra of $^{15}$NH$_3$ illustrate the potential of the line list for detections and as an efficient assistant in spectroscopic assignments. The line list is available from \url{www.exomol.com}.

\end{abstract}

\begin{keywords}
molecular data – opacity – planets and satellites: atmospheres – stars: atmospheres – ISM: molecules.
\end{keywords}

\newpage

\section{Introduction}

\NH{3}{15}  has long been detected in interstellar molecular clouds \citep{91ShWaMa.15NH3} where it is used as a tracer of the $^{15}$N/$^{14}$N isotopic ratio
\citep{02ChRoxx.15NH3,09GeMaBi.15NH3,10LiWoGe.15NH3,23ReBiCa.15NH3},  planetary atmospheres
\citep{00FoLeBe.15NH3,04FoIrPa.15NH3,14FlGrOr.15NH3} and the Earth atmosphere \citep{90HaShxx.15NH3}, in meteorites \citep{12PiWixx.15NH3}, and in comets \citep{11MuChxx.15NH3}.  Very recently, \citet{23BaMoPa.15NH3} detected \NH{3}{15} in the atmosphere of a cool brown dwarf with with the Mid-Infrared Instrument of JWST.

This work is an update of our previous empirical room temperature line lists  for \NH{3}{15} \citep{15Yurchenko} as well line lists  by \citet{11HuScLea.NH3,11HuScLeb.NH3}.
 Here we present an extended variationally computed line list for \NH{3}{15} covering the rotational excitations up to $J=30$ and the wavenumber range up to 10000~\cm\ ($\lambda \geq 1$~$\mu$m). The line list is computed using the variational program \trove\ \citep{TROVE_prog} and the same computational set up used by \citet{19CoYuTe} to produce a hot line list for \NH{3}{14} called CoYuTe. The line list is then improved using a  MARVELisation procedure, where the theoretical energies of \NH{3}{15} are replaced by experimentally-derived values obtained using the \Marvel\ procedure \citep{07FuCsTe,jt908}. To this end, an extensive set of experimental spectroscopic line positions of \NH{3}{15} has been collected from the literature \citep{47GoCo, 73ChFrxx.15NH3, 76FrOk, 77KaKrPa.15NH3, 78Jones, 80CaTrVe, 80Cohen, 80Sasada, 81SaWo.15NH3, 82LoFuTr, 82SaHaAm.15NH3, 83ShScxx.15NH3, 83UrPaBe.15NH3, 84UrDcRa.15NH3, 85DCUrRa.15NH3, 85UrDcRa.15NH3, 85UrMiRa.15NH3, 86SaScxx.15NH3, 86UrDCMa.15NH3, 87DCunha.15NH3, 90DeRaPr.15NH3, 91MoNaSh.15NH3, 92HuFrNo.15NH3, 94BuLuTa.15NH3, 94ScReFu.15NH3, 94UrKlYa.15NH3, 96BrMa, 00AnJoLe.15NH3, 06LeLiLi.15NH3, 07LiLeXu.15NH3, 08LeLiXu.15NH3, 14CeHoVe.15NH3, 16FoVaRiHe, 19CaLoFuTa, 20CaLoFu.15NH3} covering publication from 1947 (see Section~\ref{MARVEL}), combined and organised to produce a self-consistent spectroscopic network and then used to produce  experimentally derived \Marvel\ energies of \NH{3}{15}.
A new simplified and self-consistent reduced scheme to represent the ro-vibrational quantum numbers of \NH{3}{15} is suggested consisting of six quantum labels.

In order to improve more calculated energies that do not have MARVEL counterparts, the so-called isotopologue interpolation (IE) procedure is employed. This procedure takes advantage of extensive experimental spectroscopic data available for the main isotopologue of a molecule and an observation that errors of the calculated ro-vibrational energies computed the same \ai\ models are transferable between isotopologues of the same type \citep{17PoKyLo}.

The line list, comprising the ro-vibrational energies, lifetimes, Einstein coefficients, partition functions as well as  temperature and pressure dependent opacity tables is provided from the  ExoMol data base at \url{www.exomol.com}.  The line list is validated by comparing to HITRAN as well as to experimental spectra of \NH{3}{15}.

\begin{figure}
\centering
\includegraphics[width=0.95\textwidth]{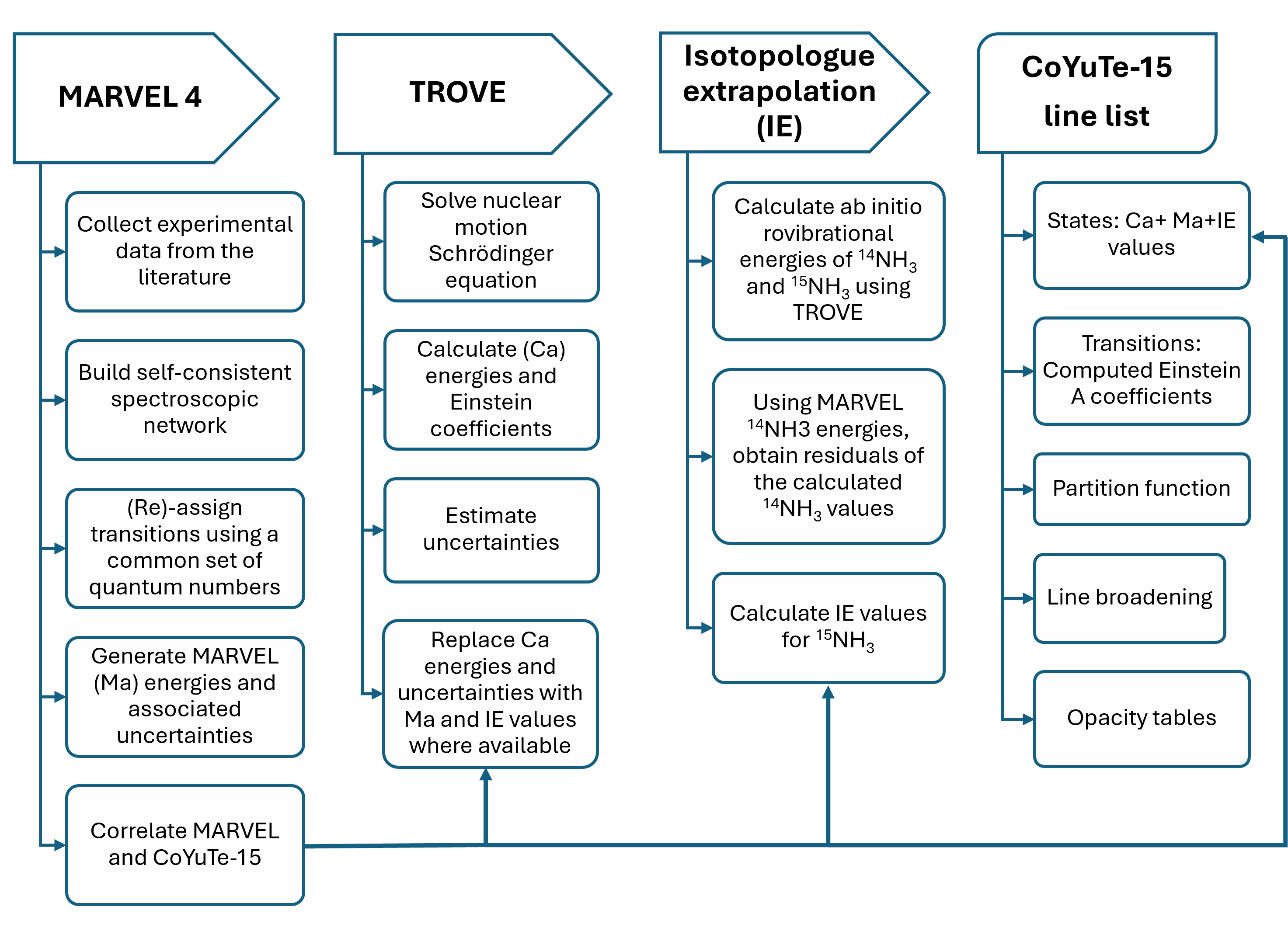}
\caption{\label{f:work-flow} Line list construction steps.}
\end{figure}

For the construction of the line list for \NH{3}{15} we followed the following steps as outlined in the work-flow diagram in Fig.~\ref{f:work-flow}: (i)  compilation of experimental line positions of \NH{3}{15} from the literature and construction of experimentally derived energies using the MARVEL procedure \citep{07FuCsTe}; (ii) Variational calculations of ro-vibrational energies and Einstein A coefficients using the program \TROVE\ \citep{TROVE_prog}; (iii) application of the isotopologue extrapolation technique and (iv) production of the line list.

\section{\Marvel\ procedure}
\label{MARVEL}

\subsection{The approach and input structure}

The algorithm used for deriving empirical energy levels of \NH{3}{15} in this work is the Measured Active Rotation Vibration Energy Level (\Marvel) algorithm, developed by Furtenbacher and co-workers \citep{07FuCsTe, 12FuCsa, 12FuCsxx.marvel, jt750,jt908}. The initial step in the \Marvel\ methodology involves compiling a dataset of measured high-resolution spectroscopic transition frequencies commonly referred to as line positions.

The line positions used as an input for \Marvel\ need to feature quantum numbers (QNs) assigned to both upper and lower states, and to ensure uniformity, a consistent set of quantum numbers is applied to all transitions. Furthermore, each line used is accompanied by its respective experimental uncertainty. The transitions are distinctly tagged, indicating the experimental source, and assigned a specific identifier within that source. Lastly, through the implementation of a weighted linear least-squares inversion procedure, \Marvel\ extracts a set of energy levels along with their corresponding uncertainties;
here these uncertainties are actually recomputed using a bootstrap procedure \citep{jt908}.

The goal is to create a cohesive network of energy levels, referred to as a spectroscopic network \citep{11CsFuxx.marvel, 07CsCzFu.marvel, 14FuArMe.marvel}. The transition uncertainties are systematically improved and incorrect lines are simultaneously removed  from the database. In more detail, a continuous assessment of the self-consistency of experimental transition frequencies is conducted. \Marvel\ identifies outliers, often caused by inaccurately assigned wavenumbers or during the transcription of the measured data into the transition list. In particular, transitions are flagged as invalid if the ratio between the optimal uncertainty and the original uncertainty exceeds a threshold of 100. These invalidated transitions are either corrected or annotated with a negative sign applied to their wavenumber values, to exclude them from further analysis by \Marvel.

Previously, the transition uncertainties were often manually increased in order to find the optimal set of uncertainties. However, the latest version of \Marvel\ offers a  bootstrap method to iteratively  determine the optimal energy level uncertainties \citep{jt908}. With the bootstrap feature of \Marvel{} 4  the individual uncertainties of the  \Marvel\ energies are determined to be suitable and reasonable. This is the method that was used in this work.

The \Marvel{} 4 Online version \citep{jt908} which was used in the current study  offers flexibility in handling transition energy units in comparison to earlier versions of \Marvel\ that were limited in only using \cm. \Marvel{} 4 Online supports various units (\cm, Hz, kHz, MHz, GHz, and THz). To enable this expanded functionality, a segment file is included as an input, containing a compilation of the experimental source names with their respective units.


\subsection{NH$_3$ quantum numbers}

For the rotation and vibration states and motion of the pyramid like ammonia molecule with a feasible inversion mode, the \Dh{3}(M) molecular symmetry group is used. This group consists of six irreducible representations, $A'_1$, $A'_2$, $E'$, $A''_1$, $A''_2$ and $E''$, referred to below  also as symmetries. As  for \NH{3}{14}, the ro-vibrational states of \NH{3}{15}  with $A'_1$ and $A''_1$ symmetries do not exist due to the Pauli principle \citep{98BuJexx}.

\subsection{Vibrational quantum numbers}

There are four vibrational modes in ammonia, $\nu_1$ (symmetric stretch), $\nu_2$ (inversion), $\nu_3$ (asymmetric stretch) and $\nu_4$ (asymmetric bend), where $\nu_3$ and $\nu_4$ are doubly degenerate.
The corresponding normal  mode vibrational quantum numbers are $n_1$, $n_2$, $n_3$, $l_3$, $n_4$, $l_4$, $l$ and s/a where $n_i \ge 0$ ($i=1,2,3,4$) are the vibrational quanta, $l_3$ and $l_4$ are the corresponding vibrational angular momentum quantum numbers with  $l = l_3 + l_4$ as the total vibrational angular momentum quantum number and  the  labels s and a are used to indicate if the inversion mode $\nu_2$ is the symmetric or  asymmetric, respectively. The values of $l_i$ ($i=3,4$) satisfy
$$
l_i = n_i,n_i-2, n_i-4,\ldots -(n_i-4), -(n_i-2), -n_i.
$$

All excitations of $\nu_1$ transform as $A'_1$ in \Dh{3}(M). The symmetric property of the symmetric bending mode $\nu_2$ upon the inversion  (symmetric or asymmetric) provides the s/a labels to the full vibrational state. Inversion state  of type $n_2\nu_2$ ($n_2=1,2,3,\ldots$) transform as $A'_1$ (s) or $A''_2$ (a) in \Dh{3}(M).
The values of $l_i$ have a direct correlation with the symmetry of the corresponding contribution, based on the simple rule of multiple of three: states with $l_i \ne  3n$ have the symmetry $E'$  forming  degenerate pairs of the same values of $L_i=|l_i|$. States with $l_i = \pm 3n$ have symmetries $A'_1$ or $A'_2$ depending on the state parity $\tau = 0$ or $\tau = 1$, respectively, see Table~\ref{t:symm:vib}.

The total vibrational symmetry $\Gamma_{\rm vib}$ is then the product of  irreducible representations $\Gamma_{i}$ from all four modes $\nu_i$, after the associated reduction to the irreducible representation. For example, the state $\nu_1+\nu_2$ ($n_1=1,n_2=1,n_3=0,n_4=0$) has the symmetry $A''_2$ as a direct symmetry product  $A'_1\otimes A''_2 \otimes A'_1 \otimes A'_1$, the state $3\nu_2 + \nu_4^{|l_4|=1}$ has the symmetry $E'$ as the product $A'_1\otimes A''_2 \otimes E' \otimes A'_1$, while a more complex example $\nu_3^{|l_3| = 1} + \nu_4^{|l_4| = 2} $ spans three states:  $(\nu_3^{1} + \nu_4^{2})^{|l|=3, A'_1}$, $(\nu_3^{1} + \nu_4^{2})^{|l|=3, A'_2}$ and $(\nu_3^{1} + \nu_4^{2})^{|l|=1, E'}$ as the result of the product \citep{98BuJe.method}
$$
 A'_1 \otimes A'_1 \otimes E' \otimes E' = A'_1 \oplus A'_2 \oplus E'.
$$
The latter can be also interpreted as a the result of the combination of the bending and stretching vibrational angular momenta $l_3 = \pm 1$ and $l_4 = \pm 2$, respectively:
\begin{gather}
l = l_3+ l_4 = \left\{
\begin{array}{ccc}
     +1+2 & = +3 & A'_1/A'_2  \\
     -1-2 & = -3 & A'_1/A'_2  \\
     +1-2& =  -1 & E' \\
     -1+2& =   1 & E' \\
    \end{array}
\right.
\end{gather}
Here, $l$ as a total vibrational angular momentum is important   to resolve the degeneracy of the three distinct components  of the combinations $l_3,l_4$ in the $\nu_3^{|l_3| = 1} + \nu_4^{|l_4| = 2} $, which is necessary for the for the non-degenerate $l=\pm 3$ states. The same result,  in principle,  is achieved by using the total vibrational symmetry $\Gamma_{\rm vib}$, with $A^\dagger_1$ and $A^\dagger_2$ ($\dagger='$ or $''$) to account for the two non degenerate components of this combination. In fact, using symmetries  is more physically sensible  than the non-physical alternative with the defined signs  $l=3,-3$, but the former is more popular in experimental studies of ammonia and  spherical tops in general.

\begin{table}

\caption{Correlation between the vibrational quantum numbers and \Dh{3}(M) symmetries of ammonia ($n=1,2,3,\ldots$, $m=1,2,3,\ldots$, $\dagger=', ''$). Here $\dagger=\prime$\ if  $K$ is even, and $\dagger=\prime \prime$ when $K$ is odd.
\label{t:symm:vib}}
{\centering
    \begin{tabular}{ccccccccccccccc}
         \cline{1-3} \cline{5-7} \cline{9-11} \cline{13-15}
    \multicolumn{3}{c}{$n\nu_1$} & & \multicolumn{3}{c}{$n\nu_2$} && \multicolumn{3}{c}{$n\nu_3^{l_3}$ or  $n\nu_4^{l_4}$} &&  \multicolumn{3}{c}{$\ket{J,K,\tau_{\rm rot}}$} \\
        \cline{1-3} \cline{5-7} \cline{9-11} \cline{13-15}
         $n$ &  & $\Gamma$ && a/s &  & $\Gamma$ && $l_i$ & $\tau$ & $\Gamma$ && $K$ & $\tau_{\rm rot}$  & $\Gamma_{\rm rot}$ \\
           \cline{1-3} \cline{5-7} \cline{9-11} \cline{13-15}
         any    &  0 &  $A'_1$ &&  s    &   &  $A'_1$ &&  0    &  0 &  $A'_1$ && $3n $ & 0 & $A^\dagger_1$  \\
         & & &&         a     &   &  $A''_2$ &&  $3m$   &  0 &  $A'_1$ && $3n $ & 1 & $A^\dagger_2$  \\
         & & && & & && $3m$   &
         1 &  $A'_2$ && $3n\pm 1 $ & 0,1  & $E^\dagger$  \\
          & & && & & &&  $3m\pm 1$ &   &    $E'$ \\
     \cline{1-3} \cline{5-7} \cline{9-11} \cline{13-15}
    \end{tabular}
    }
\end{table}

\subsection{Rotational quantum numbers}

The rotational assignment follows the rigid rotor wavefunctions  $\ket{J,k,m}$, where $J$ is the total angular momentum quantum number, $m$ is the projection of the total angular momentum on the laboratory fixed axis $Z$ and $k$ is the  projection of the total angular momentum on the molecular fixed axis $z$. Since $m$ does not play any role in the classification of the rotational states in the absence of external fields, it is commonly omitted from the consideration.

In order to correlate to the molecular symmetry, the rigid rotor wavefunctions  $\vert J,k,m\rangle$ are first combined into symmetric and asymmetric Wang-type functions $\ket{J,K,\tau_{\rm rot}}$ \citep{05YuCaJe}
\begin{eqnarray}
   \label{e:|J0tau>}
   \ket{J,0,\tau_{\rm rot}},  & \tau_{\rm rot} = J\;{\rm mod}\; 2, \\
   \label{e:|Jtau>}
\ket{J,K,\tau_{\rm rot}} &= \frac{1}{\sqrt{2}} \left[ \ket{J,|k|} +  (-1)^{J+K+\tau_{\rm rot}}\ket{J,-|k|} \right] , \quad \tau_{\rm rot} = 0,1,
\end{eqnarray}
where we introduced $K=|k|$ and $\tau_{\rm rot}$ to describe the parity of the rotational state and dropped $m$. The symmetries associated with different values of $K$ and $\tau_{\rm rot}$ are explained  in Table~\ref{t:symm:vib}. Although just two quantities,  $J$ and $K=|k|$ are sufficient for the description of the degenerate $E^{\dagger}$ states, when $K=3n$,  one more descriptor is required. In experimental studies, it is usually the sign of $k$ that serves this purpose, which is however not well defied and often lead to ambiguity in descriptions of the experimental data from  different sources. Here we use the symmetries $\Gamma_{\rm rot}$ =  $A^\dagger_1$ and $A^\dagger_2$ to distinguish different states of the same $|K|=3n$. This symmetry label is also used for  states with $3n\pm1$ when $\Gamma_{\rm rot} = E^\dagger$ for consistency.


\subsection{Ro-vibrational quantum numbers}

It should be noted that the absolute signs of $l_3$, $l_4$ or $l$ are also ill-defined. Indeed,  the positive and negative values of $l$ (and also $k$, see below) contribute equally to the description of  ro-vibrational states of symmetric tops giving rise to the so-called  Wang 50/50 mixtures, see Eq.~\eqref{e:|Jtau>}. The relative signs however do matter, which is the main reason for  the signed values of $l$s commonly used in the spectroscopic analysis.
Another reason is that the sign of $l$ or $k$ is effectively an additional quantum label allowing to distinguish states of  the same absolute $|l|$ or $|k|$, even though they often  lead to conflicting assignments of the same spectra from different experimental sources.
Ammonia is characterised by its tunneling motion which leads to inversion of the molecule and splitting of the levels into symmetric/antisymmetric combinations which we denote s/a. Depending on the choice of QNs, s/a can be redundant but still useful as the inversion splitting is a key motif of ammonia spectra.

Our quantum numbers of choice are  the following (unambiguous) quantities: $J$, $n_1$, $n_2$, $n_3$, $n_4$,  $L_3=|l_3|$, $L_4=|l_4|$, $L=|l_3+l_4|$ (i.e. absolute values of the corresponding vibrational angular momenta), $K=|k|$ (absolute value of the rotational quantum number), irreducible representations $\Gamma$, $\Gamma_{\rm vib}$ and $\Gamma_{\rm rot}$ of the total ro-vibrational state, as well as of the  vibrational and rotational contributions and s/a. These quantum numbers correspond closely to the ones recommended by \citet{jt546} and used in the previous
\Marvel\ studies of \NH{3}{14} \citep{jt608,20FuCoTe}.


It is now becoming more and more common to use  simplified, compact  schemes for assigning ro-vibrational states where several quantum numbers are combined into a single counting index. For example, in  recent experimental studies of CH$_4$ (see \citet{jt926}), it is a counting ro-vibrational index within the same $J$, total symmetry $\Gamma$ and the polyad $P$. In the case of N$_2$O \citep{jt908}, this is the index also within the same $J$, $l$ (vibrational angular momentum) and polyad $P$. As will be discussed below, the vibrational quantum numbers of ammonia become ambiguous at high vibrational excitations as they are coupled through rotational interactions, which mix vibrational states from the same and different polyads. Moreover, the coupling scheme between different degrees of freedom are complicated by the symmetry product rules making the usage of the vibrational quantum number less intuitive. We therefore decided to introduce a reduced QN scheme, with the minimal  number of  quantum indices that  maintain the key aspects of the rotation-vibrational states. One of such aspects we would like to preserve is the description of the individual vibrational ($J=0$) and rotational state contributions we would like to preserve. While the rotational part in the case of the symmetric top ammonia can be mainly described by the rotational quantum numbers $J$ and $K$,  as a compact representation of the full set  of the vibrational quantum numbers   $n_1,n_2,n_3^{l_3},n_4^{l_4},l,\Gamma_{\rm vib}$, we introduce  a single vibrational index $N$. This index counts vibrational s/a pairs if states sorted according to their increased $J=0$ energies:
\begin{equation}
\label{e:exp:QN}
N = \left\{ n_1,n_2,n_3^{l_3},n_4^{l_4},l,\Gamma_{\rm vib} \right\}.
\end{equation}
Note that it does not contain s/a, i.e. the full inversion-vibrational description is according with $N^{\rm s/a}$.
Thus, our final set of the ro-vibrational quantum numbers  consists of the following six labels given by:
\begin{equation}
\label{e:MARVEL:QN}
{\rm QN} = \left\{ J, \Gamma, N, s/a, K, \Gamma_{\rm rot} \right\},
\end{equation}
where we included the rotational symmetry $ \Gamma_{\rm rot}$ in order to resolve the ambiguity occurring at $K=3n$  for the non-degenerate $A_1^{\dagger}/A_2^{\dagger}$ components. While we replace the full vibrational description by $\{ N, {\rm s/a}\}$ in the \Marvel\ set for the sake of its compactness, the original eight-QN information is retained by providing a detailed correlation table between $N$ and $\left\{ n_1,n_2,n_3^{l_3},n_4^{l_4},l,\Gamma_{\rm vib} \right\}$, which is given below in Section~\ref{s:MARVEL}.


\section{Selection rules}

The standard rigorous (electric dipole, single photon) selection rules in the case of ammonia spectra are
\begin{align}
&\Delta J = 0, \pm 1  \\
& J'+J'' \ne 0, \\
& A'_2 \leftrightarrow A''_2 \quad ({\rm ortho}), \\
& E' \leftrightarrow E'' \quad ({\rm para}),
\end{align}
where $A'_1$ and $A''_1$ do not exist due to the Pauli principle (the corresponding nuclear statistical weights are zero). Since the  parity ($'$ or $''$) changes through a electric dipole, single photon transition, which is also directly correlated with the s/a characteristics, one can also obtain  the following rigorous rules for s/a:
$$
\begin{array}{ll}
  \Delta K\;   {\rm even}:& {\rm s} \leftrightarrow {\rm a} \\
  \Delta K \;  {\rm odd}:&  {\rm s} \leftrightarrow {\rm s}, \quad {\rm a} \leftrightarrow {\rm a}.
\end{array}
$$
This is because $K$ flips  the parity of the vibrational states when $K$ is odd. For the two-photon transitions, the selection rules are
\begin{align}
&\Delta J = 0, \pm 1, \pm 2  \\
& J'+J'' \ne 0, \\
& A^\dagger_2 \leftrightarrow A^\dagger_2 , \\
& E^\dagger \leftrightarrow E^\dagger.
\end{align}
A number of two-photon spectra have been recorded experimentally \citep{76FrOk,78Jones,83ShScxx.15NH3,00AnJoLe.15NH3}, but no transitions with $\Delta J = \pm 2$ have been observed; these transitions are allowed but expected to be too weak to
readily observable \citep{83ShScxx.15NH3}.

Most of the experimental data collected in this work are from  single-photon electric dipole spectra.
There are no transitions between states of para and ortho states and therefore a single estimated ortho-para transition  frequency (magic number) was required to connect the ortho- and para- networks (see below).

Ammonia has a rather strong $\Delta K =0$ propensity rule which can cause networks to fragment and problems with effective Hamiltonian representations, see \citet{jt784} for a discussion of this issue. For the current dataset, half  of transitions (49\%) are with $\Delta K =0$, 37\% with $\Delta K =1$, 8\% with $\Delta K =2$  and only 6\% is for all other transitions $\Delta K>2$.


\section{Experimental data sources}
\label{sec:marvel_source}

In the following, the existing experimental sources containing transition frequencies of \NH{3}{15} are reviewed. Detailed statistics, including the number of lines considered and validated, are presented in Table~\ref{t:sources:MARVEL}.

\textbf{47GoCo} \citep{47GoCo} Twenty millimeter (kHz) wave inversion transitions $Q(J)$ with uncertainty 0.02~kHz. 

\textbf{73ChFr} \citep{73ChFrxx.15NH3} A millimeter wave rotation-inversion line R(1) in the   $\nu_2$  hot band with uncertainty 0.5 MHz; quantum numbers were taken from 83UrPaBeKr. 

\textbf{76FrOk}  \citep{76FrOk} Two-photon (forbidden) transitions in  the $\nu_{2}$ band. Uncertainty provided.

\textbf{77KaKrPa} \citep{77KaKrPa.15NH3} One inversion line, $\nu_2-\nu_2$ hot band. 

\textbf{78Jones} \citep{78Jones} Eleven infrared  transitions of the $\nu_2$ band; a/s labels of IR for some transitions had to be swapped to match other data.  78Jones also report two-photon transitions of \NH{3}{15} with incomplete (or confusing) assignment which we could not match to the data from other sources and excluded.

\textbf{80CaTrVe} \citep{80CaTrVe} Rotation-inversion spectrum; the stated uncertainties are 0.01 \cm\ for blended lines or 0.001 \cm\ otherwise.

\textbf{80Cohen}  \citep{80Cohen} Inversion band within $\nu_4$;  

\textbf{80Sasada} \citep{80Sasada} Microwave inversion band.  Uncertainty of 0.1 MHz was assumed.



\textbf{81SaWo}   \citep{81SaWo.15NH3} The $\nu_2$ band.
Here it was important to include the vibrational symmetry into the state assignment for states with $K=3$ due to the degeneracy associated with states of $A_1$/$A_2$. We used \TROVE\  quantum numbers and energies for reconstructing the state symmetries for all states.

\textbf{82DiFuTrMi} \citep{82LoFuTr} The  $\nu_2$, $2\nu_2$, $3\nu_2$, $\nu_4$, and $\nu_2$ + $\nu_4$ bands.
Stated uncertainty  = 0.0001 \cm, which we doubled  for blended lines.  1883 transitions were validated out of 1926. Some $l$'s were changed  to match more recent experimental assignment (19CaLoFaTa) if the  upper state energies in the associated combination differences (CDs) matched. The upper values of ``s/a'' were assigned using the rule of ($K'-K''$) should odd/even for  ``s/a'' to change/not to change.

\textbf{82SaHaAmSh}  \citep{82SaHaAm.15NH3} The   $\nu_4$, 2$\nu_2$ bands. Some of the  $2\nu_2/\nu_4$ assignments were swapped when matching to other sources.
The uncertainties were set to 0.05~\cm, but the data appear to be of lower quality and inconsistent with  this uncertainty; they conflict with many other observations leading to large differences in CDs, significantly higher this the uncertainty. Only 385 out of 775 transitions could be validated.

\textbf{83ShSc} \citep{83ShScxx.15NH3} The $\nu_2$ band, 120 two-photon forbidden transitions; uncertainty was stated in the paper of 0.0002~\cm.

\textbf{83UrPaBeKr}: \citep{83UrPaBe.15NH3} The $\nu_2$ and $\nu_2$ -- $\nu_2$ bands.	
All negative $k$ were changed to positive $K=|k|$ to make consistent with out convention. 
The uncertainty values are stated as 0.001 --  0.00003 \cm.



\textbf{84UrDCRa}  \citep{84UrDcRa.15NH3} Five forbidden transitions in the $\nu_2$ band.

\textbf{85DcUrRa} \citep{85DCUrRa.15NH3} The $\nu_2$ band. Uncertanties are provided.

\textbf{85UrDCRaPa}:  \citep{85UrDcRa.15NH3}   Bands $2\nu_2$ and $\nu_4$.
Accurate  ground state (g.s.) energies and $\nu_2$ energies are reported, which were them used in the combination differences calculations. We changed some $l$s to match more recent data of \citet{19CaLoFuTa} based on the CDs (i.e. on the match of the upper state energies). Some of the $2\nu_2$ states were reassigned to $\nu_4$ using \TROVE\ matches and to agree with CDs from other sources. Many of the uncertanties were increased to 0.1 \cm\ based on the CDs analysis.
Only 8 transitions could not be validated out of 807.



\textbf{85UrMiRa} \citep{85UrMiRa.15NH3} The overtone $\nu_1+\nu_2$  and hot $\nu_1+\nu_2-\nu_2$ bands.
Uncertanties are provided.

\textbf{86SaSc} \citep{86SaScxx.15NH3}  Hot band $2\nu_2$--$\nu_2$.  Uncertanties are provided.


\textbf{86UrDCMa} \citep{86UrDCMa.15NH3} The $\nu_2$ band. The upper state s/a were defined using the usual rule for the single-photon transitions. Uncertanties are provided.

\textbf{87DCunha}  \citep{87DCunha.15NH3}  hot band $2\nu_2$--$\nu_2$. Uncertainties are provided and claimed to be better than 0.002~\cm.

\textbf{90DeRaPrUr} \citep{90DeRaPr.15NH3} The $\nu_4$ band.  Mainly  intensities are reported alongside line frequencies. It is not clear if the line frequencies are experimental or theoretical. Excluded.

\textbf{91MoNakShi} \citep{91MoNaSh.15NH3}	Reports transitions in the region of the $5\nu_1$ band but  no specific sub-band  assignment is given.
The assignments are very limited and we could not make them work so this dataset was   excluded.


\textbf{92HuFrNo}  \citep{92HuFrNo.15NH3} The 26 hot band $\nu_4-\nu_4$ transitions; uncertainties  0.02-0.05 MHz as provided.

\textbf{94BuLuTaMa} \citep{94BuLuTa.15NH3} Five pure rotational lines in kHz.


\textbf{94ScReFuIz}: \citep{94ScReFu.15NH3} The  $\nu_2-\nu_2$ hot band. Far infrared emission lines, produced by both inversion (single-photon) and Raman (two-photon) processes. Large uncertainties of 0.05~\cm\ were assumed due to 2 decimal places given.  Some are pure rotational within g.s. and some are hot within $\nu_2$. Not all lines could be validated.


\textbf{94UrKlYa} \citep{94UrKlYa.15NH3}	Far-infrared ground state rotation-inversion transitions. Uncertainty  is stated to be 0.000003 \cm.

\textbf{96BrMa} \citep{96BrMa} 5294 \cm\ region containing  $\nu_3+\nu_4$ band. All $l'=1$/$l'=-1$ (non-physical for $\nu_3+\nu_4$)  were reassigned to $l'=2$/$l'=-2$.
96BrMa.22 excluded because it disagrees with 96BrMa.18.  Reported uncertainty  0.0003~\cm\ in the paper.

\textbf{00AnJoLeSa} and \textbf{00AnJoLeSa(IR)}  form \citep{00AnJoLe.15NH3}. \textbf{00AnJoLeSa} contains  $\nu_1$  Raman (two-photon forbidden) transitions;   00AnJoLeSa.72 is from a  nonphysical state ($J=0$, $A_1$ g.s.) and was excluded.
\textbf{00AnJoLeSa(IR)} contains  single photon (IR) transitions. We separated these transitions into two parts to facilitate handling these two distinct data sets.

\textbf{06LeLiLiXu} \citep{06LeLiLi.15NH3}  The  $\nu_1+\nu_3$ band. Some transitions were reassigned to $2\nu_1$ and $(\nu_3+2\nu_4)^{1}$ to match \TROVE.

\textbf{07LiLeXu} \citep{07LiLeXu.15NH3}. The  $\nu_1+2\nu_4$ band, transitions and term values.

\textbf{08LeLiXu} \citep{08LeLiXu.15NH3} Reported transitions from the 1.5 $\mu$m region containing the $\nu_3+2\nu_4$  band.	Some lines were  changed to $\nu_1+\nu_3$, $2\nu_1$, $\nu_1+2\nu_4$ based on the CDs analysis.

\textbf{14CeHoVeCa}  \citep{14CeHoVe.15NH3}   Bands $\nu_1$, $\nu_2+\nu_3$, $2\nu_2+\nu_3-\nu_2$ from the 2.3 $\mu$m region. All but four transitions were validated (179 out of 183).

\textbf{16FoVaRiHe}	\citep{16FoVaRiHe}. Transitions from the $\nu_1+\nu_3$ band are reported but without specific assignment of the individual bands, i.e. all transitions are initially assigned to $\nu_1+\nu_3$. Some of the lines were vibrationally reassigned by a combined CD analysis with the assigned data from 08LeLiXu \citep{08LeLiXu.15NH3} covering the same upper states. Many lines even with multiple CDs could not be validated  and were  excluded as they lead to upper energies which conflict with those given by other sources and our \TROVE\ calculations.



\textbf{19CaLoFuTa(MHz)} and \textbf{19CaLoFuTa},  from  \citep{19CaLoFuTa}.  \textbf{19CaLoFuTa(MHz)} contains $\nu_4-\nu_4$ data in MHz, while \textbf{19CaLoFuTa}  reports transitions from the $\nu_2-\nu_2$, $2\nu_2-2\nu_2$ and $\nu_4-\nu_4$ bands are in \cm. We separate 19CaLoFuTa (MHz) from  19CaLoFuTa (\cm) because of the structure of the \Marvel4 input requiring the data in the original units. Only 10 transitions out of 7563 could not be validated.



\textbf{20CaLoFu} \citep{20CaLoFu.15NH3}. This a second largest source reporting 5770 transitions from the bands $\nu_2+\nu_4$ and $3\nu_2$. About 250 transitions could not be validated because of conflicts with the experimental data from other sources as well as unusually large differences with  our theoretical  (\TROVE) estimates. Among those cases, the most unexpected disagreement was found for five $\nu_2+\nu_4$ energy levels of 20CaLoFu ($J=5,6,7,8$), all supported by multiple  (up to 9) transitions with CDs within about 0.02 \cm, that conflicted with the corresponding theoretical energies of \TROVE\ by more than  8--16 \cm. Considering that in all other cases, \TROVE\ energies agree with \Marvel\ to much better than 1~\cm, these 5 groups of  transitions were first considered as suspicious and then treated as experimental outliers, possibly caused by impurities present in experimental spectra of \citet{20CaLoFu.15NH3}.  Moreover, in all these case, we could find at least transition from 82SaHaAmSh from the same levels that agreed with \TROVE\ within 0.01~\cm. An example of such suspected false CDs is illustrated in Table~\ref{t:20CaLoFu}, where we show 8 transitions from 20CaLoFu with a common upper state $J=5, K'=1, a, A'_2$, $\Gamma_{\rm rot = } E''$ and $N = $ 5 ($\nu_2+\nu_4$). The corresponding combination  seemingly agree with each other leading to an upper energy term value of 2874.127~\cm. This prediction is however very far from the \TROVE\ value of 2882.4683~\cm, which we does not seem to be unrealistic considering the generally very good agreement for most of the  \Marvel\ energies (94\% are within 0.1~\cm), see Fig.~\ref{f:obs-clac}, including the $\nu_2+\nu_4$ band. Moreover, the experimental data set  82DiFuTrMi \citep{82LoFuTr} also contains a transition to this state with the upper state energy of 2882.4596~\cm, which is an excellent agreement with the \TROVE\ prediction, see Table~\ref{t:20CaLoFu}. We have encountered at least 4 similar  cases with a large number of CDs ($\sim$8), all for $\nu_2+\nu_4$ with $J=5,6,7,8$ from 20CaLoFu \citep{20CaLoFu.15NH3}, with large differences from \TROVE\ (8--16~\cm) and in every case with an alternative, single transition from   82DiFuTrMi \citep{82LoFuTr} in an excellent agreement with \TROVE. Although this is not obvious, but we decided to trust the \TROVE\ and 82DiFuTrMi  predictions in those cases and excluded the corresponding transitions from \Marvel.


\begin{table}
    \centering
    \caption{Example of combination differences from 20CaLoFu \citep{20CaLoFu.15NH3} conflicted with the \TROVE\ prediction of the upper state energy of 2882.4683~\cm\ as well as of the value from the
 single line 82DiFuTrMi \citep{82LoFuTr}, 2882.459601~\cm, which agree with each other. }
    \label{t:20CaLoFu}
    {
    \scriptsize
    \begin{tabular}{rrrrrccrlrr}
    \hline\hline
     \multicolumn{1}{c}{$\tilde{\nu}$, \cm} &\multicolumn{1}{c}{unc, \cm}& $J''$ & $K''$  & $a/s''$ & $\Gamma''$ & $\Gamma''_{\rm rot}$ & $N$& \multicolumn{1}{c}{line} & \multicolumn{1}{c}{$\tilde{E'}$, \cm} & \multicolumn{1}{c}{$\tilde{E''}$, \cm} \\
    \hline
      134.74608  &    0.0004 &$    4$&$  1$&$ s $&$A''_2$&$ E''  $&$ 6 $& 20CaLoFu.112       &  2874.127322  &   2739.38124   \\
     1046.35599  &    0.0060 &$    4$&$  1$&$ s $&$A''_2$&$ E''  $&$ 4 $& 20CaLoFu.2818      &  2874.127558  &   1827.77157   \\
      809.85126  &    0.0006 &$    4$&$  0$&$ a $&$A''_2$&$ A'_1 $&$ 3 $& 20CaLoFu.2006      &  2874.127853  &   2064.27659   \\
     1497.88856  &    0.0006 &$    6$&$  0$&$ a $&$A''_2$&$ A'_1 $&$ 2 $& 20CaLoFu.3924      &  2874.128003  &   1376.23944   \\
     2458.29564  &    0.0006 &$    6$&$  0$&$ a $&$A''_2$&$ A'_1 $&$ 1 $& 20CaLoFu.5059      &  2874.128241  &    415.83260   \\
       597.3536  &    0.0006 &$    6$&$  0$&$ a $&$A''_2$&$ A'_1 $&$ 3 $& 20CaLoFu.1254      &  2874.128864  &   2276.77526   \\
      952.45873  &    0.0006 &$    5$&$  1$&$ s $&$A''_2$&$ E''  $&$ 4 $& 20CaLoFu.2585      &  2874.129251  &   1921.67052   \\
     1714.09336  &    0.0006 &$    4$&$  0$&$ a $&$A''_2$&$ A'_1 $&$ 2 $& 20CaLoFu.4453      &  2874.129451  &   1160.03609   \\
       2466.627  &    0.0040 &$    6$&$  0$&$ a $&$A''_2$&$ A'_1 $&$ 1 $& 82DiFuTrMi.1633    &  2882.459601  &    415.83260   \\
\hline\hline
       \end{tabular}
    }
\end{table}


\subsection{\Marvel}
\label{s:MARVEL}

By the \Marvel\ definition, \Marvel\ energy of the lowest state is set to zero, which in case of \NH{3}{15} is $J=0$, $K=0$, a, $A''_2$ (0,0,0$^0$,0$^0$), i.e. the upper, asymmetric inversion component of the ground state state, with the symmetric  component $J=0$, $K=0$, s, $A'_2$ (0,0,0$^0$,0$^0$) non-existent due to the nuclear spin statistics. Because of this effect, the inversion splitting of ammonia \NH{3}{15} cannot be observed experimentally and is not present in our \Marvel\ set.

\begin{figure}
\centering
\includegraphics[width=0.7\textwidth]{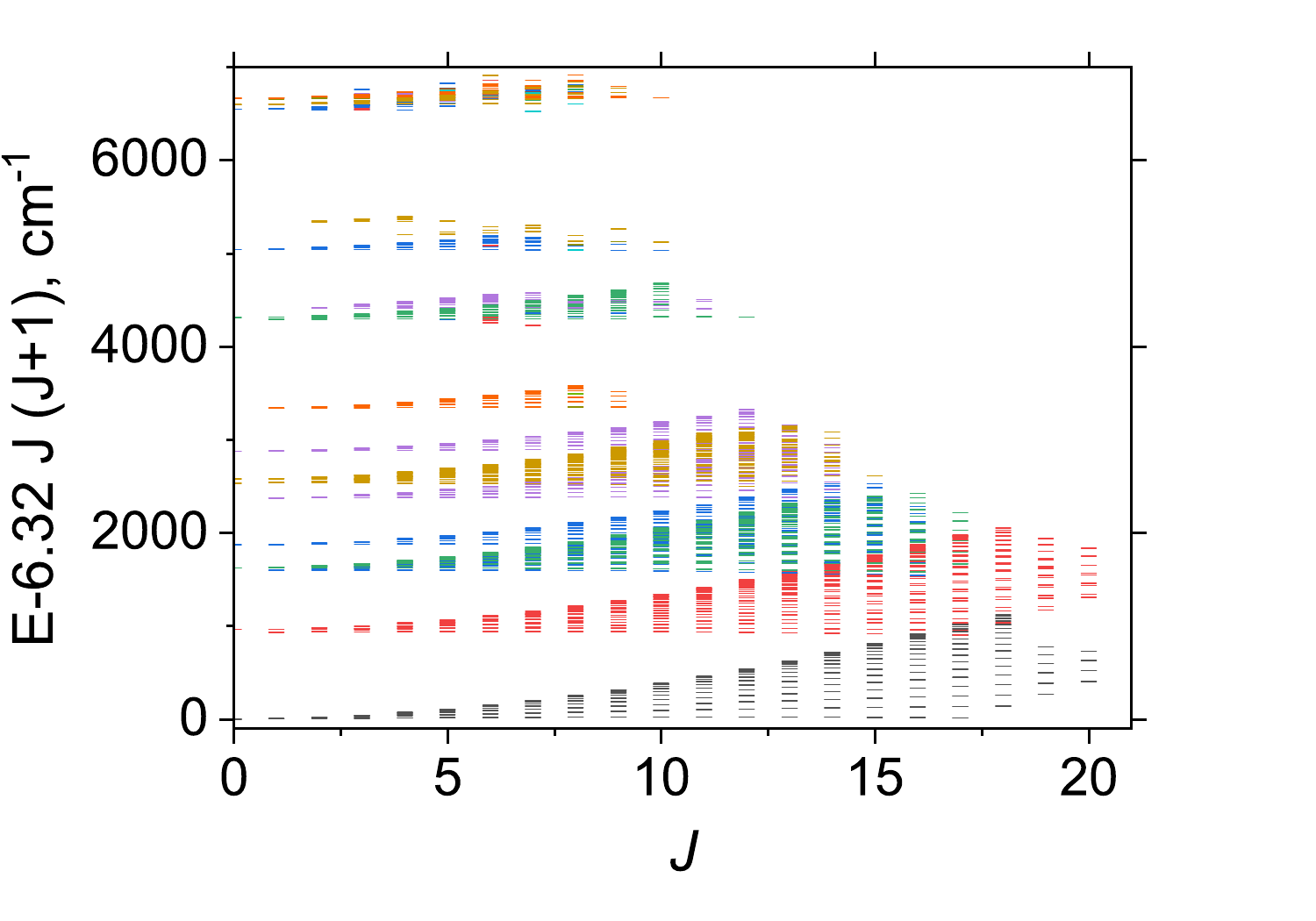}
\caption{\label{f:MARVEL-E} \Marvel\ reduced energy term values of $^{15}$NH$_3$. Different colours are used to indicate different vibrational bands.}
\end{figure}

In order to  connect the ortho- and para- networks,   we used a MAGIC number as an energy difference between lowest (ground state) energy levels of  para- and ortho- states, $J=1$, $K=1$, s ($E''$)  and  $J=0$, $K=0$, a ($A''_2$), respectively,  taken from the term values list by \citet{85UrDcRa.15NH3}: $\Delta \tilde{E}$ = 15.391430~\cm.
Figure~\ref{f:MARVEL-E} shows the \Marvel\ energies of \NH{3}{15}.

\begin{table}
\caption{Description of the \Marvel\ data of the experimental transitions of  \NH{3}{15} used in this work. A/V are the numbers of the available and validated transitions, respectively; unc is  mean uncertainty, in the units of the original source; the label s/d indicate  single/double photon transitions. }
    \label{t:sources:MARVEL}
    {
\scriptsize
\centering
\begin{tabular}{lcr@{$-$}lcrcp{5cm}}
\hline\hline
Source   &    Units  &    \multicolumn{2}{c}{Range}  &   A/V &    unc   &  s/d &  Bands            \\
\hline
      47GoCo       &  MHz  &      20272.04 &     25323.51 &  0.02000  &        20/20       &    s   &  rot-inv  \\
      73ChFr       &  MHz  &     175054.90 &    175054.90 &  0.50000  &         1/1        &    s   &  rot-inv   \\
      76FrOk       &  \cm  &        921.75 &       987.63 &  0.00020  &        19/19       &    d   &  $\nu_2-\nu_2$  \\
     77KaKrPa      &  MHz  &     430038.30 &    430038.30 &  0.20000  &         1/1        &    s   &  $\nu_2-\nu_2$\\
     78Jones       &  \cm  &        888.04 &      1090.97 &  0.00020  &        11/11       &   s/d  &  $\nu_2$   \\
     80CaTrVe      &  \cm  &         38.92 &       276.87 &  0.00319  &       195/194      &    s   &  rot-inv \\
     80Cohen       &  MHz  &      21904.29 &     55581.98 &  0.02009  &        45/41       &    s   &  $\nu_4-\nu_4$  \\
     80Sasada      &  MHz  &       8180.56 &     44160.66 &  0.09922  &       115/114      &    s   &  rot-inv        \\
      81SaWo       &  \cm  &        926.84 &      1082.12 &  0.00020  &        14/14       &    s   &  $\nu_2$        \\
    82DiFuTrMi     &  \cm  &        511.11 &     13017.41 &  0.00575  &      1926/1853     &    s   &  $\nu_2$, $2\nu_2$, $3\nu_2$, $\nu_4$, $\nu_2$ + $\nu_4$       \\
    82SaHaAmSh     &  \cm  &        697.00 &      1838.71 &  0.05000  &       775/385      &    s   &  $\nu_4$, 2$\nu_2$                                             \\
      83ShSc       &  \cm  &        916.75 &      1091.54 &  0.00020  &       120/118      &    d   &  $\nu_2$        \\
    83UrPaBeKr     &  \cm  &         18.31 &      1098.02 &  0.00061  &       207/205      &    s   &                   $\nu_2$, $\nu_2$ -- $\nu_2$                  \\
     84UrDCRa      &  \cm  &        809.44 &      1026.49 &  0.00200  &         5/5        &    s   &  $\nu_2$        \\
     85DCUrRa      &  \cm  &        697.74 &      1225.72 &  0.00050  &       499/499      &    s   &  $\nu_2$        \\
    85UrDCRaPa     &  \cm  &       1129.21 &      2100.01 &  0.00493  &       807/791      &    s   &                        $2\nu_2$, $\nu_4$                       \\
     85UrMiRa      &  \cm  &       3163.21 &      4495.55 &  0.00922  &       484/483      &    s   &               $\nu_1+\nu_2$,  $\nu_1+\nu_2-\nu_2$              \\
      86SaSc       &  \cm  &        720.71 &      1169.02 &  0.01092  &       150/150      &    s   &  $2\nu_2$--$\nu_2$                                             \\
     86UrDCMa      &  \cm  &        533.67 &      1328.74 &  0.00505  &       433/411      &    s   &  $\nu_2$        \\
     87DCunha      &  \cm  &        695.26 &      1179.20 &  0.00260  &       199/196      &    s   &                         $2\nu_2$--$\nu_2$                      \\
    90DeRaPrUr     &  \cm  &       1494.62 &      1533.29 &  0.02000  &        33/0        &    s   &                             $\nu_4$                            \\
     91MoNaSh      &  \cm  &      15237.32 &     15535.90 &  0.00500  &        129/0       &    s   &  $5\nu_1$       \\
     92HuFrNo      &  MHz  &      30590.99 &     54321.94 &  0.02927  &        26/25       &    s   &  $\nu_4-\nu_4$   \\
    94BuLuTaMa     &  KHz  &   18871541.30 &  23421979.50 &  1.70000  &         5/5        &    s   &  rot-inv        \\
    94ScReFuIz     &  \cm  &         33.14 &      1085.15 &  0.05150  &        60/38       &    s   &  $\nu_2-\nu_2$  \\
     94UrKlYa      &  \cm  &         19.08 &       119.53 &  0.00007  &        33/33       &    s   &  rot-inv        \\
      96BrMa       &  \cm  &       4944.50 &      5179.79 &  0.00030  &       129/124      &    s   &   $\nu_3+\nu_4$   \\
    00AnJoLeSa     &  \cm  &       3328.20 &      3333.51 &  0.00233  &        56/55       &    s   &    $\nu_1$                            \\
  00AnJoLeSa(IR)   &  \cm  &       3327.62 &      3334.11 &  0.00006  &        71/71       &    d   &      $\nu_1$                            \\
    06LeLiLiXu     &  \cm  &       6372.64 &      6777.71 &  0.02000  &       272/268      &    s   &  $\nu_1+\nu_3$  \\
     07LiLeXu      &  \cm  &       6412.16 &      6687.72 &  0.00381  &        96/95       &    s   &   $\nu_1+2\nu_4$                        \\
     08LeLiXu      &  \cm  &       6446.10 &      6817.69 &  0.04387  &       224/219      &    s   &     $\nu_3+2\nu_4$, $\nu_1+\nu_3$, $2\nu_1$, $\nu_1+2\nu_4$    \\
    14CeHoVeCa     &  \cm  &       4275.01 &      4339.22 &  0.00200  &       183/179      &    s   &          $\nu_1$, $\nu_2+\nu_3$, $2\nu_2+\nu_3$-$\nu_2$        \\
    16FoVaRiHe     &  \cm  &       6300.39 &      6790.79 &  0.00600  &       590/522      &    s   &    $\nu_1+\nu_3$                         \\
    19CaLoFuTa     &  \cm  &          5.84 &      2126.90 &  0.00076  &      7449/7428     &    s   &    $\nu_4-\nu_4$                         \\
 19CaLoFuTa(MHz)   &  MHz  &      21904.29 &    168831.50 &  0.02380  &        69/66       &    s   &    $\nu_2-\nu_2$,$2\nu_2-2\nu_2$,$\nu_4-\nu_4$ (and between)   \\
     20CaLoFu      &  \cm  &         67.06 &      3076.22 &  0.00111  &      5770/5447     &    s   &   $\nu_2+\nu_4$, $3\nu_2$                    \\
      MAGIC        &  \cm  &         15.39 &      2175.89 &   0.00010 &         2/2        &    s   &  rot-inv       \\

\hline
\hline
\end{tabular}
}
\end{table}

\subsection{\Marvel\ and correlation between experiment and theory}


The experimental transitions in our list all originate from six lower states, g.s., $\nu_2$, $\nu_4$,  $2\nu_2$, $2\nu_4$ and $\nu_2+\nu_4$, which correspond to $N=$ 1, 2, 3, 4, 5 and 6, respectively.  We started by building  combination differences for experimental transitions originating from g.s. and $\nu_2$, where we could take advantage of the existing energy term values from \citet{85UrDcRa.15NH3}. By applying these term values to the transitions from
\citep{82LoFuTr,82SaHaAm.15NH3,85UrDcRa.15NH3,86SaScxx.15NH3,87DCunha.15NH3,90DeRaPr.15NH3,19CaLoFuTa,20CaLoFu.15NH3}, the combination differences for the upper state  $\nu_4$, $2\nu_2$, $2\nu_4$ and $\nu_2+\nu_4$  could be constructed and the corresponding energies compared. This procedure helped identify and correct  possible misassignments of many upper states. In turn, the corresponding term values  could be used to construct combination differences for all other hot bands originated from these states.

To facilitate this procedure, the \name\ line list was  extensively used. Our ultimate goal is to use the experimentally derived  \Marvel\ energies in place of the theoretical values where available. To this end, the \Marvel\ energies need to be correlated to the calculated \TROVE\ energies. Being able to compare the two sets of data to each other helps both ways, to establish the quality of the variational calculations but also resolve potential problems in  the \Marvel\ set.

Owing to the high quality of the underlying empirical potential energy surface of NH$_3$, the theoretical ro-vibrational term values and associated combination differences  were of reasonably high quality, see Fig.~\ref{f:obs-clac}. To help match experiment and theory, the signs of different  values of  $l$ and $k$, and sometimes of $l_3$ and $l_4$ in the experimental sources had to be reassigned to make them consistent between different experimental data. In many case, the associated (non-rigorous) quantum number had to be modified, including the vibrational ones, $n_1$, $n_2$, $n_3$ and $n_4$. However, the rigorous quantum numbers $J$ and symmetry were not altered. In order to correlate to \TROVE, some \TROVE\ quantum had to be modified to make them unambiguous and consistent with the experiment.

As part of the process, a set of term values was compiled to be used as lower states for the experimental set of \NH{3}{15} we collected based on a mixture of the experimentally derived term values of 85UrCrPa (g.s. and $\nu_2$) and \TROVE\ term values. We then were able to correlate all the upper and lower states from the experimental transitions to the \TROVE\ ro-vibrational states and thus use the \TROVE\ QNs as an independent self-consistent set of assignments. This procedure helped correlate the experimental transitions from different sources featuring the same states, correct assignments, identify and  exclude outliers.

Once the full correlation was constructed, it was then possible  to replace the experimental quantum numbers with our \Marvel\ set of six quantum numbers in Eq.~\eqref{e:MARVEL:QN}. This set is based on the collective vibrational quantum number $N$, which is defined as a counting index of the vibrational ($J=0$, s) energy term values, including the non-existing  $A'_1$  and $A''_1$ states.  The latter are also important as they play the role of the `band centres' for the ro-vibrational term values or transition frequencies.
The correlation between $N$ and the  vibrational quantum numbers $\left\{ n_1,n_2,n_3^{l_3},n_4^{l_4},l,\Gamma_{\rm vib} \right\}$ is illustrated in Table~\ref{t:vibindex} for all vibrational states from our \Marvel\ data. The full correlation table is given as supplementary material. We used the theoretical, \name, term values to establish the correlation between the vibration-inversion states and $N^{\rm s/a}$. Being theoretical, our quantum label $N^{\rm s/a}$ has the caveat  of depending on the order of the \name\ energies and is thus affected by its quality, especially at higher excitations.

\begin{table}
    \caption{Vibrational ($J=0$) energy term values $\tilde{E}$, experimental quantum numbers ($n_1$, $n_2$, $n_3^{l_3}$, $n_4^{l_4}$, $l$) and their mapping to the labels $N^{\rm s/a}$ and $\Gamma_{\rm vib}$  for all vibrational state used in our \Marvel\ set for \NH{3}{15}.}
    \label{t:vibindex}
\centering
\scriptsize
\begin{tabular}{rrrcrrrrrrrr}
\hline\hline
$N$    &  $\tilde{E}^{(s)}$, \cm & $\tilde{E}^{(a)}$, \cm &  $n_1$ &  $n_2$ &  $n_3$ &  $l_3$ &  $n_4$ & $l_4$ &  $l$ & $\Gamma_{\rm vib}$ \\
\hline
    1 &      0.0000 &    0.759258   &   0 &   0 &   0 &   0 &   0 &   0 &   0 &  $A'_1$     \\
    2 &    928.4200 &   962.884440  &   0 &   1 &   0 &   0 &   0 &   0 &   0 &  $A'_1$     \\
    3 &   1591.1451 &  1870.839271  &   0 &   2 &   0 &   0 &   0 &   0 &   0 &  $A'_1$     \\
    4 &   1623.1601 &  1624.223193  &   0 &   0 &   0 &   0 &   1 &   1 &   1 &  $E'$       \\
    5 &   2369.3052 &  2876.114293  &   0 &   3 &   0 &   0 &   0 &   0 &   0 &  $A'_1$     \\
    6 &   2533.3162 &  2577.600578  &   0 &   1 &   0 &   0 &   1 &   1 &  -1 &  $E'$       \\
    9 &   3233.9370 &               &   1 &   0 &   0 &   0 &   0 &   0 &   0 &  $E'$       \\
   10 &   3333.3021 &  3334.268304  &   1 &   0 &   0 &   0 &   0 &   0 &   0 &  $A'_1$     \\
   12 &   3438.6186 &               &   1 &   0 &   0 &   0 &   0 &   0 &   0 &  $A'_1$     \\
   14 &             &  4161.664679  &   0 &   1 &   0 &   0 &   2 &   0 &   0 &  $A''_2$    \\
   15 &   4125.4594 &  4181.129616  &   0 &   1 &   0 &   0 &   2 &   2 &   2 &  $E'$       \\
   16 &   4288.0489 &  4312.285044  &   1 &   1 &   0 &   0 &   0 &   0 &   0 &  $A'_1$     \\
   17 &   4403.6103 &  4421.482245  &   0 &   1 &   1 &   1 &   0 &   0 &   1 &  $E'$       \\
   19 &   4742.9544 &               &   0 &   0 &   1 &   1 &   1 &   1 &   2 &  $A'_1$     \\
   20 &             &  5095.012196  &   0 &   0 &   1 &   1 &   1 &   1 &   2 &  $E''$      \\
   21 &   4790.9936 &  4793.138030  &   0 &   0 &   1 &   1 &   1 &   1 &   2 &  $E'$       \\
   26 &   5040.5521 &  5041.175789  &   0 &   0 &   1 &   1 &   1 &   1 &   2 &  $A'_2$     \\
   27 &   5041.2057 &  5041.819217  &   0 &   0 &   1 &   1 &   1 &   1 &   2 &  $E'$       \\
   30 &   5129.3266 &  5333.267022  &   0 &   2 &   1 &   1 &   0 &   0 &   1 &  $E'$       \\
   41 &             &  6702.902353  &   0 &   0 &   1 &  -1 &   2 &   2 &   1 &  $A''_2$    \\
   43 &   6294.3854 &  6660.407789  &   1 &   0 &   0 &   0 &   2 &   2 &   2 &  $E'$       \\
   45 &             &  6693.154085  &   0 &   0 &   2 &   0 &   3 &   3 &   1 &  $A''_2$    \\
   48 &             &  6364.423717  &   0 &   0 &   0 &   0 &   4 &   4 &   2 &  $E''$      \\
   50 &   6511.6301 &  6513.176308  &   2 &   0 &   0 &   0 &   0 &   0 &   0 &  $A'_1$     \\
   51 &   6546.9982 &  6548.468141  &   1 &   0 &   0 &   0 &   2 &   2 &   2 &  $E'$       \\
   52 &   6570.4264 &               &   1 &   0 &   2 &   0 &   0 &   1 &   1 &  $E'$       \\
   53 &   6595.0789 &  6595.961735  &   2 &   0 &   0 &   0 &   0 &   0 &   0 &  $A'_1$     \\
   54 &   6596.5904 &  6597.461983  &   1 &   0 &   1 &   1 &   0 &   0 &   1 &  $E'$       \\
   55 &   6636.6412 &               &   0 &   0 &   1 &  -1 &   2 &   2 &   1 &  $A'_2$     \\
   56 &   6637.6712 &  6638.904945  &   0 &   0 &   1 &   1 &   2 &   2 &   1 &  $A'_1$     \\
   57 &   6651.4442 &  6650.418625  &   0 &   0 &   2 &   2 &   1 &   1 &   1 &  $E''$      \\
   58 &   6664.6067 &  6665.278000  &   0 &   0 &   1 &  -1 &   2 &   2 &   1 &  $E'$       \\
   59 &   6681.2824 &               &   0 &   0 &   1 &   1 &   2 &  -2 &  -1 &  $A'_1$     \\
   60 &   6692.8086 &               &   1 &   0 &   1 &   1 &   0 &   0 &   1 &  $A'_2$     \\
   61 &   6698.1314 &               &   0 &   0 &   1 &   1 &   2 &  -2 &  -1 &  $E'$       \\
   62 &   6705.3820 &               &   1 &   0 &   1 &   1 &   0 &   0 &   1 &  $E'$       \\
   63 &   6709.1670 &               &   0 &   2 &   1 &   1 &   1 &   1 &   1 &  $A'_1$     \\
   66 &   6833.9506 &  6834.678719  &   1 &   0 &   1 &   1 &   0 &   0 &   1 &  $E'$       \\
\hline
\hline
\end{tabular}

\end{table}

As an illustration of the \Marvel\ data structure, in Table~\ref{t:MARVEL:trans} we provide an extract from the main \Marvel\ data set. In order to help the users with our compact  assignment scheme, especially with the  new  counting index $N^{\rm s/a}$, in the supplementary material we also provide (i) the original \name\ vibrational  term values   $N^{\rm s/a}$ is based on, accompanied by the mapping with the experimental quantum numbers from Eq.~\eqref{e:exp:QN} (see Table~\ref{t:vibindex}); (ii) the experimental transitions using the native quantum numbers (subject to some corrections for internal consistency).

\begin{table}
\caption{Extract from the \Marvel\ transition file. The \Marvel\ frequency wavenumber $\tilde{\nu}$ and uncertainties are in cm$^{-1}$. The mapping between the compact vibrational index $N$ and the full set of the vibrational quantum numbers, see Eq.~\eqref{e:exp:QN}, is illustrated in Table~\ref{t:vibindex}. There are two uncertainty columns to allow the input uncertainty to be updated while retaining the original uncertainty of the source. }
    \label{t:MARVEL:trans}
\scriptsize
\centering
\begin{adjustbox}{max width=1\textwidth,center}
    \begin{tabular}{d{6}d{6}d{6}rrrrrrrrr rrrrrrrr l}
    \toprule
    \toprule
    \multicolumn{1}{c}{ $\tilde{\nu}$} & \multicolumn{1}{r}{\rm unc., (\cm)} &\multicolumn{1}{r}{\rm unc., (\cm)} &  \multicolumn{6}{c}{\rm Quantum `numbers' of upper states}  &
                                                 \multicolumn{6}{c}{\rm Quantum `numbers' of lower states}  & {\rm Source }   \\
&& & $J'$ &$K'$ & $s/a'$ & $\Gamma'$  &  $\Gamma'_{\rm rot}$ & $N'$
& $J''$ &$K''$ & $s/a''$ & $\Gamma''$  &  $\Gamma''_{\rm rot}$ & $N''$  \\
    \midrule
   5158.7121 &    0.0003 &    0.0003 &    6 &    1 &  s    &  A2"  &  E"   &    27 &     5 &     0 &  s    &  A2'  &  A2'  &     1 &  96BrMa.128 \\
   5179.7856 &    0.0003 &    0.0003 &    7 &    1 &  a    &  A2'  &  E"   &    27 &     6 &     0 &  a    &  A2"  &  A1'  &     1 &  96BrMa.129 \\
   19.083628 &  0.000003 &  0.000003 &    1 &    0 &  s    &  A2'  &  A2'  &     1 &     0 &     0 &  a    &  A2"  &  A1'  &     1 &  94UrKlYa.1 \\
   38.913812 &   0.00006 &   0.00006 &    2 &    1 &  s    &  E"   &  E"   &     1 &     1 &     1 &  a    &  E'   &  E"   &     1 &  94UrKlYa.2 \\
   40.390961 &   0.00004 &   0.00004 &    2 &    0 &  a    &  A2"  &  A1'  &     1 &     1 &     0 &  s    &  A2'  &  A2'  &     1 &  94UrKlYa.3 \\
   40.403877 &   0.00004 &   0.00004 &    2 &    1 &  a    &  E'   &  E"   &     1 &     1 &     1 &  s    &  E"   &  E"   &     1 &  94UrKlYa.4 \\
   58.713871 &   0.00004 &   0.00004 &    3 &    0 &  s    &  A2'  &  A2'  &     1 &     2 &     0 &  a    &  A2"  &  A1'  &     1 &  94UrKlYa.5 \\
     58.7166 &   0.00006 &   0.00006 &    3 &    1 &  s    &  E"   &  E"   &     1 &     2 &     1 &  a    &  E'   &  E"   &     1 &  94UrKlYa.6 \\
    \toprule
    \toprule
    \end{tabular}
    \end{adjustbox}
\end{table}

\section{Line list production}

\subsection{\TROVE\ calculations}

The \name\ line list has been produced using the  variational code \trove\ \citep{TROVE_prog} with the spectroscopic model from \citet{19CoYuTe} developed and used for the CoYuTe line list for \NH{3}{14}, where we switched to the (atomic) masses  $m_{\rm N} = $ $15.000108898$~Da and $m_{H} = 1.007825032$~Da. This spectroscopic model consists of an empirical potential energy surface (PES) fitted by \citet{18CoOvPo} and an \ai\ electric dipole moment surface (DMS) from \citet{09YuBaYa.NH3}. For the details of the computational set up see \citet{19CoYuTe}, which we only briefly describe here.

The kinetic energy operator (KEO) was expanded in terms of five linearised coordinates $\xi_{i}^{\rm lin}$, $i=1\ldots 5$ around a non-rigid (inversion) frame on a equidistant grid of 1000 points covering the umbrella coordinate $\delta$ from -55~to 55$^{\circ}$. The umbrella coordinate is defined as an angle between any of the N-H$_i$ modes and their trisector  (see \citet{23Yurchenko}). The five  coordinates $\xi_{i}^{\rm lin}$ are constructed from the following valence coordinates: the three bond lengths $r_i$ and the asymmetric combinations $S_{a}$ and $S_{b}$ of the bond angles $\alpha_{i}$ ($i=1,2,3$) defined as
\begin{align}
    S_a &= \frac{1}{\sqrt{6}} \left[  2\alpha_{1} - \alpha_{2} - \alpha_3 \right], \\
    S_b &= \frac{1}{\sqrt{2}} \left[ \alpha_{2} - \alpha_3 \right]
\end{align}
by expanding them  in terms of  the Cartesian displacements around the reference configuration and truncating after the first order, see \citet{TROVE_prog}.

The stretching basis functions $\phi_{v_1}(\xi_1^{\rm lin})$, $\phi_{v_2}(\xi_2^{\rm lin})$ and $\phi_{v_3}(\xi_3^{\rm lin})$, as well as the inversion basis functions $\phi_{v_6}(\xi)$ ($\xi = \delta$) are  generated numerically using the Numerov-Cooley approach \citep{24Numerov.method,61Cooley.method}, while  for the  bending basis functions $\phi_{v_4}(\xi_4^{\rm lin})$ and $\phi_{v_5}(\xi_5^{\rm lin})$, 1D Harmonic oscillator wavefunctions were used. These basis functions are then optimised via a contraction-symmetrisation procedure \citep{17YuYaOv} to form a symmetry adapted (\Dh{3}(M))  vibrational ($J=0$) basis functions $\Phi_{N}^{\Gamma_{\rm vib}}$ as a sum-of-products of $\phi_{v_i}$ ($i=1,\ldots,6$). At this stage, the vibrational basis functions $\Phi_{N}^{\Gamma_{\rm vib}}$ are assigned the normal mode quantum numbers $n_1$, $n_2$, $n_3^{l_3}$, $n_4^{l_4}$ by correlating our local node wavefunctions $\phi_{v_i}$  to the Harmonic-oscillator-like solutions. For the rotational basis set, symmetrised  rigid-rotor wavefunctions $\ket{J,K,\Gamma_{\rm rot}}$ as simple Wang combinations, see Eqs.(\ref{e:|J0tau>}, \ref{e:|Jtau>}).  The rotational and vibrational assignments are then propagated to the final ro-vibrational variational wavefunctions using the largest eigen-coefficient approach \citep{23Yurchenko}. The size of the vibrational basis set is controlled by the polyad number parameter $P$ defined as follows:
\begin{equation}
P = 2 (v_1 + v_2 + v_3) + v_4 + v_5 + v_6 = 2 (n_1+n_3) + n_2 + n_4 \le P_{\rm max}.
\end{equation}
As in \citet{19CoYuTe}, $P_{\rm max} = 17$ was used.  Another parameter controlling the ro-vibrational basis set size is the  energy threshold $E_{\rm max} = 30000$~\cm.
For the  Zero-Point-Energy defined at the energy of the (non-physical) $J=0, K=0$, s, $N=1$ state relative to the minimum of the CoYuTe PES NH$_3$  \citep{19CoYuTe} we obtained  8001.42~\cm.

\begin{figure}
\centering
\includegraphics[width=0.7\textwidth]{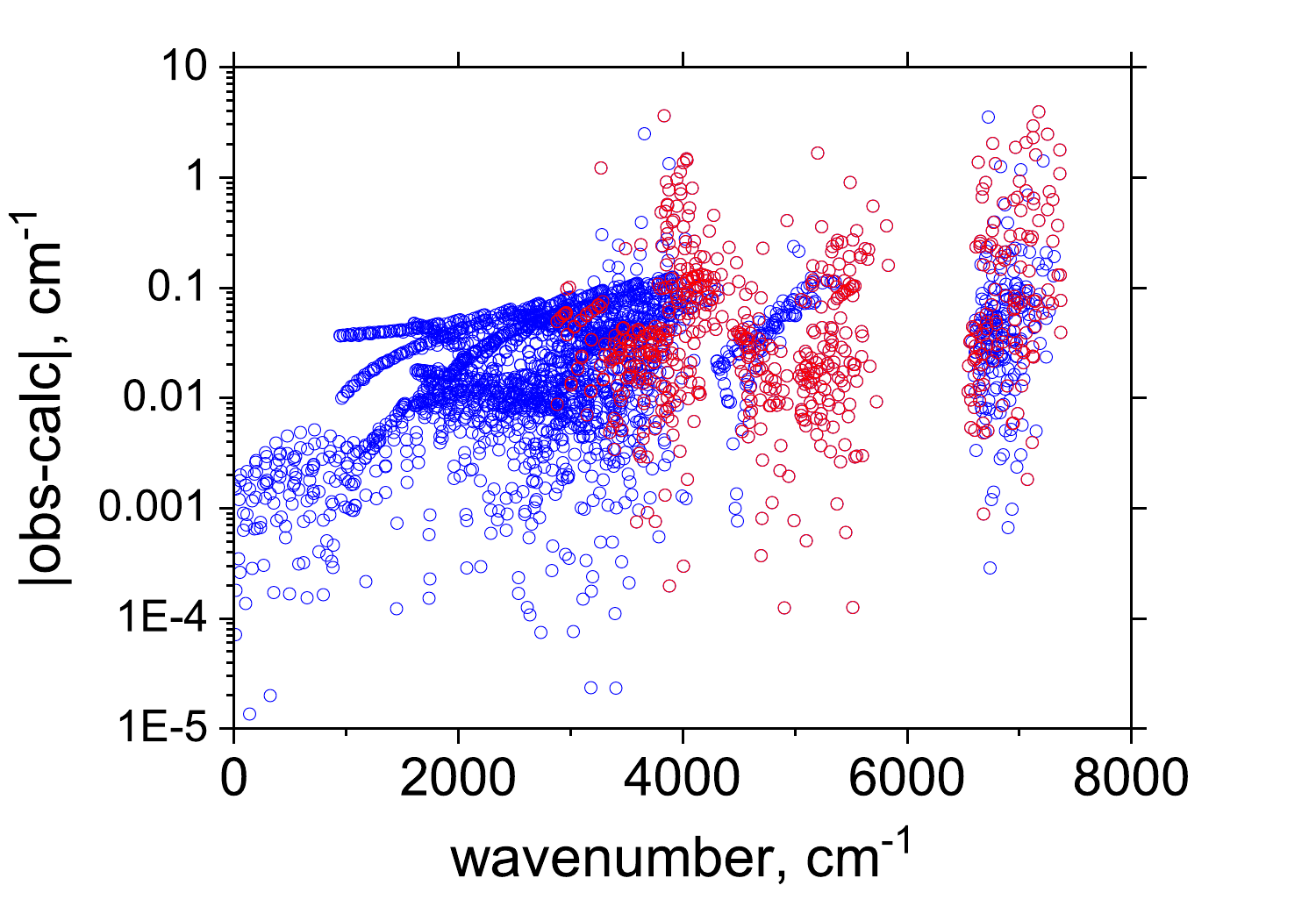}
\caption{\label{f:obs-clac} Obs.-Calc. (\Marvel-\name) as a function of the energy term value. The red colour indicates \Marvel\ energies defined by one or single transitions only.}
\end{figure}

Figure~\ref{f:obs-clac} shows how the calculated energy values with \TROVE\ compare to the experimentally derived energy term values of \NH{3}{15} (\Marvel) as obs.-calc. residuals.  The majority of the residuals are below 0.1~\cm. The majority of the values with large residuals correspond to the \Marvel\ energies based on 1--2 transitions only.

\subsection{Pseudo-\Marvel: isotopologue energy extrapolation}

One of the limitations of the \Marvel\ technique is that the completeness of the \Marvel\ energies depends on the completeness of the original experimental sources. It is not uncommon for the experimental data set to be patchy, with some lines missing, with ambiguous assignment or blended transitions. In attempt to  fill gaps in the \Marvel\ energy sets when MARVELising the line lists, different techniques can be employed \citep{jt948}. Here we consider the isotopologue energy level extrapolation technique, similar to the one employed in \citet{17PoKyLo}, where we leverage the present \Marvel\ energy levels list for the \NH{3}{14} parent isotopologue \citep{20FuCoTe}. The isotopologue extrapolation (IE) technique uses the assumption that the calculation errors of ro-vibrational energies of different isotopologues based on the same Born-Oppenheimer PES are approximately uniform.

To this end, variational \TROVE\ calculations are performed using the \ai\ surface from \citet{16PoOvKy}, for both the \NH{3}{14} and \NH{3}{15} isotopologues. The \ai\  ro-vibrational energies from both isotopologues are then matched to their corresponding \Marvel\ energies from this study and on the parent isotopologue from \citet{20FuCoTe}. Here we opt for the \ai\ PESs in place of (more accurate) empirical PESs in order to mitigate possible contamination of the Born-Oppenheimer original \ai\ surface with any mass dependent effects absorbed into CoYuTe PES through fitting to the experiment in \citet{19CoYuTe}. The \ai\ residuals $\Delta \tilde{E} = \tilde{E}^{\rm obs.}-\tilde{E}^{\rm calc.}$ (observed minus calculated)  of each energy level included in \Marvel\ are first computed for the main isotopologue \NH{3}{14} and then these obs.-calc. differences are added onto the \ai\  energies of \NH{3}{15}. In this way we form isotope extrapolation (IE) \citep{jt948} \Marvel\ levels for \NH{3}{15} enabling a semi-empirical determination of energy levels present in the \NH{3}{14} \Marvel\ list but not yet determined for \NH{3}{15}.

\begin{figure}
    \centering
\includegraphics[width=0.6\textwidth]{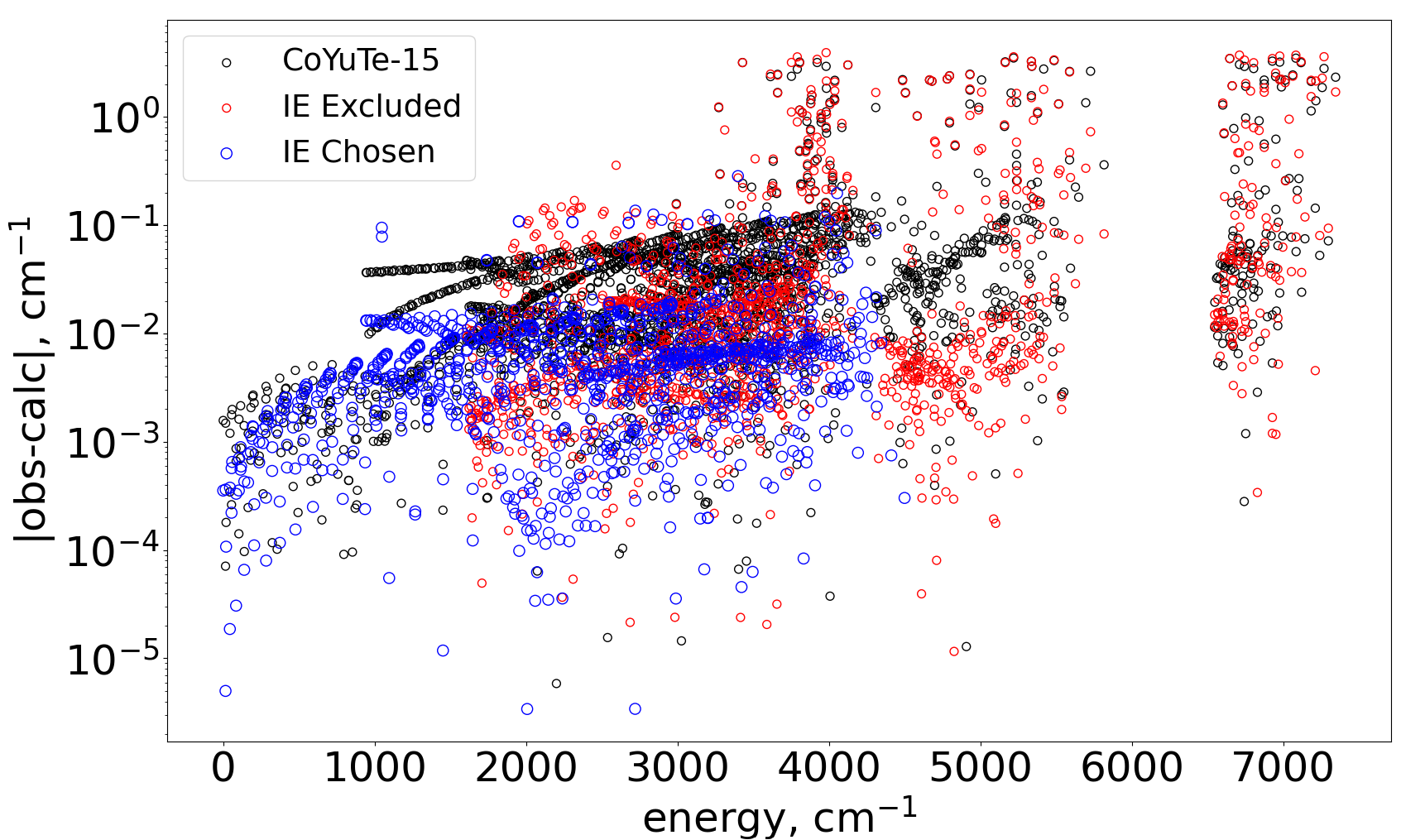}
    \caption{ Absolute values of residuals  $| \tilde{E}^{\rm obs.}-\tilde{E}^{\rm calc.}|$ ($|$obs-calc$|$) between the \name\ energies and the  \Marvel\ energies alongside the IE residuals  as a function of the energy term value. IE levels from bands selected for substitution are highlighted separately.}
    \label{f:CoYuTeAndIEResiduals}
\end{figure}

To gauge the accuracy of the IE levels for \NH{3}{15}, we compare those levels present in our \Marvel\ study, with their IE counterparts. The residuals  between the \Marvel\ energies \NH{3}{15} and the \name\ computed energies alongside the residuals of IE energies are given in Fig.~\ref{f:CoYuTeAndIEResiduals}. As can be seen, the difference in accuracy between the two is found to vary, and shows a vibrational dependence. For this reason, we only substitute IE levels in place of \name\ levels for bands which are known to provide increased accuracy over \name. For excited vibrational states, the only IE levels found to improve over \name\ belong to the ground state, the $\nu_2$ inversion mode and its overtones, $2\nu_2$ and  $3\nu_2$, summarised in Table~\ref{t:IESubstitutedBands}, and indicated in Fig.~\ref{f:CoYuTeAndIEResiduals} as `IE Chosen'.  The corresponding uncertainties, also listed in Table~\ref{t:IESubstitutedBands}, were estimated by comparing to the \Marvel\ values if available,  which are indicated in Fig.~\ref{f:CoYuTeAndIEResiduals} as `IE Excluded'.


\begin{table*}
\caption{IE and \name\  root-mean-squared obs.-calc. of $^{15}$NH$_3$ bands chosen for IE substitution.}
\begin{tabular}{rrrrr}
    \toprule\toprule
    $N$ & $a/s$ & State  &  \multicolumn{1}{c}{IE (cm$^{-1}$)} &  \multicolumn{1}{c}{\name\ (cm$^{-1}$)}\\
    \midrule
   1  & s & g.s.    &  0.019505 & 0.021751 \\
   1  & a & g.s.    &  0.018846 & 0.019437 \\
   2  & s & $\nu_2$ &  0.023396 & 0.064547 \\
   2  & a & $\nu_2$ &  0.022278 & 0.065662 \\
   4  & a & $2\nu_2$&  0.027318 & 0.059184 \\
   5  & s & $3\nu_2$&  0.018277 & 0.035313 \\
   6  & a & $3\nu_2$&  0.024460 & 0.058929 \\
   \bottomrule\bottomrule
\end{tabular}
\label{t:IEUncertainty}
\end{table*}

\begin{table*}
\caption{Uncertainties  $\delta_{\text{IE}}$ used for IE levels based on the state, $J$ and the uncertainty $\delta_{14}$ of the corresponding \Marvel\ level for the $^{14}$NH$_3$ isotopologue.}
\label{t:IESubstitutedBands}
\begin{tabular}{rrrrr}
    \toprule\toprule
    State & $a/s$ & $J$  & $\delta_{14}$ (cm$^{-1}$) &  $\delta_{\text{IE}}$ (cm$^{-1}$)\\
    \midrule
    g.s.  & s & $\leq 10$  & $\leq 1.2\times10^{-3}$  &   $10^{-2}$ \\
      &  & $> 10$  & $\leq 1.2\times10^{-3}$  &   $10^{-1}$ \\
      &  & $> 10$  & $> 1.2\times10^{-3}$  &   $2\times 10^{-1}$ \\
    g.s.  & a & $\leq 10$  & $\leq 10^{-3}$  &   $10^{-2}$ \\
      &  & $> 10$  & $\leq 10^{-3}$  &   $10^{-1}$ \\
      &  & $> 10$  & $> 10^{-3}$ &   $2\times 10^{-1}$ \\
    $\nu_2$ &  s & N/A  & $\leq 1.2\times10^{-3}$  &   $10^{-1}$ \\
      &  & N/A  & $> 1.2\times10^{-3}$  &   $2\times 10^{-1}$ \\
    $\nu_2$ &  a & $\leq 10$  & $\leq 10^{-3}$  &   $10^{-2}$ \\
    &   & $> 10$  & $\leq 10^{-3}$  &   $10^{-1}$ \\
      &  & $> 10$  & $> 10^{-3}$  &   $2\times 10^{-1}$ \\
    $2\nu_2$ &  a & $\leq 10$  & N/A  &   $10^{-2}$ \\
      &  & $>10$  & N/A  &   $10^{-1}$ \\
    $3\nu_2$ &  s & N/A  & $\leq 10^{-3}$  &   $10^{-1}$ \\
      &  & N/A  & $> 10^{-3}$  &   $2\times 10^{-1}$ \\
    $3\nu_2$ &  a & N/A  & $\leq 7.5\times10^{-4}$  &   $10^{-1}$ \\
      &  & N/A  & $> 7.5\times10^{-4}$  &   $2\times 10^{-1}$ \\
   \bottomrule\bottomrule
\end{tabular}
\end{table*}

\subsection{Line list}

The \name\ line list for \NH{3}{15} was computed for $J=0\ldots 30$ covering the wavenumber range up to 10000~\cm, with the upper/lower energies limited by 18000/8000~\cm, respectively. Intensity (Einstein A coefficients) calculations  were performed on a GPU cluster using the GAIN MPI code \citep{17AlYuTe}.

The line list contains  929~795~249 transitions between 12~699~617 ro-vibrational states and consists of a  States file, Transitions files and Partition function file. According with the ExoMolHD standards \citep{jt810}, we use  experimentally derived energies to replace the theoretical values, where available, the procedure commonly referenced to as `MARVELisation'. From the \Marvel\ set of \hl{2777} energies, we selected    \hl{2754} values  with residuals smaller than 1~\cm\ from the \TROVE\ energy term value to avoid any accidental \Marvel\ outliers. These were complemented by \hl{326} IE values  and finally with a fourteen $J=18$ Effective Hamiltonian term values from \citet{85UrDcRa.15NH3}.

Extracts from the States and Transition files are illustrated in Tables~\ref{t:states} and \ref{t:trans}. We follow the energy convention used for ammonia, where the lowest energy is the energy of the non-physical ground state  $J=0$, $K=0$, s, $A'_1$ (0,0,0$^0$,0$^0$), which is set to zero. In this convention, the lowest observable state $J=0$, $K=0$, a, $A''_2$ (0,0,0$^0$,0$^0$) has the energy of the inversion splitting, which we set to  $hc\cdot$\hl{0.757685}~\cm\ \citep{85UrDcRa.15NH3}. Therefore the \Marvel\ energies, which are based  on a  different contention for the lowest ro-vibrational energy, are shifted by 0.757685~\cm\ before using them in the MARVELisation procedure. For practical purposes, we keep the non-existing $J=0$ energies in the .states file, which are however `switched-off' from any intensity applications by setting the corresponding  degeneracy factors $g$ to zero.

\begin{figure}
\centering
\includegraphics[width=0.7\textwidth]{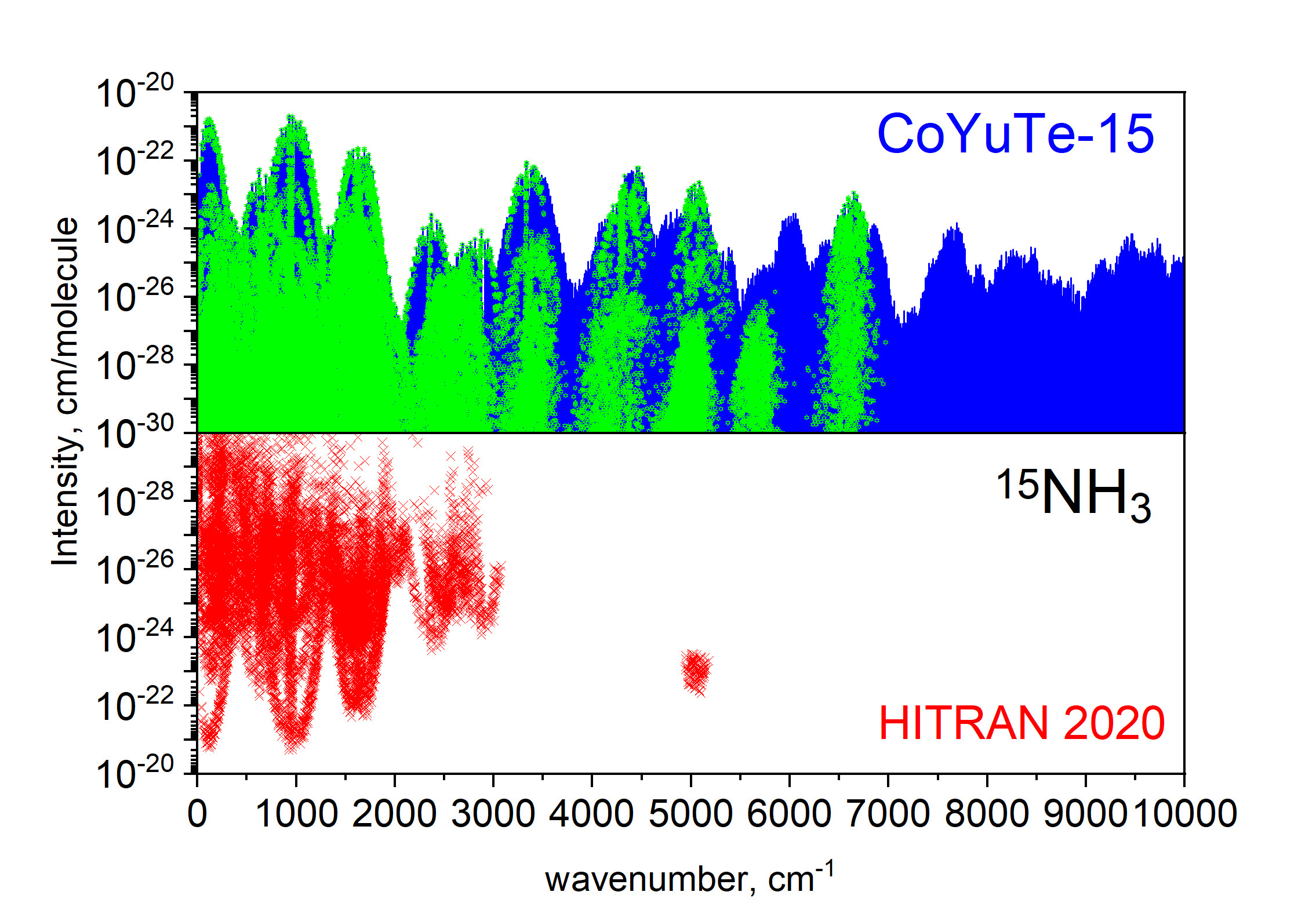}
\caption{\label{f:HITRAN} Comparison of the \name\ spectrum (scaled to natural abundance) with HITRAN~2000 \citep{HITRAN2020} at $T=296$~K. Green points indicate \hl{53~856} transitions connecting \Marvel\ states only.}
\end{figure}
\begin{figure}
\centering
\includegraphics[width=0.33\textwidth]{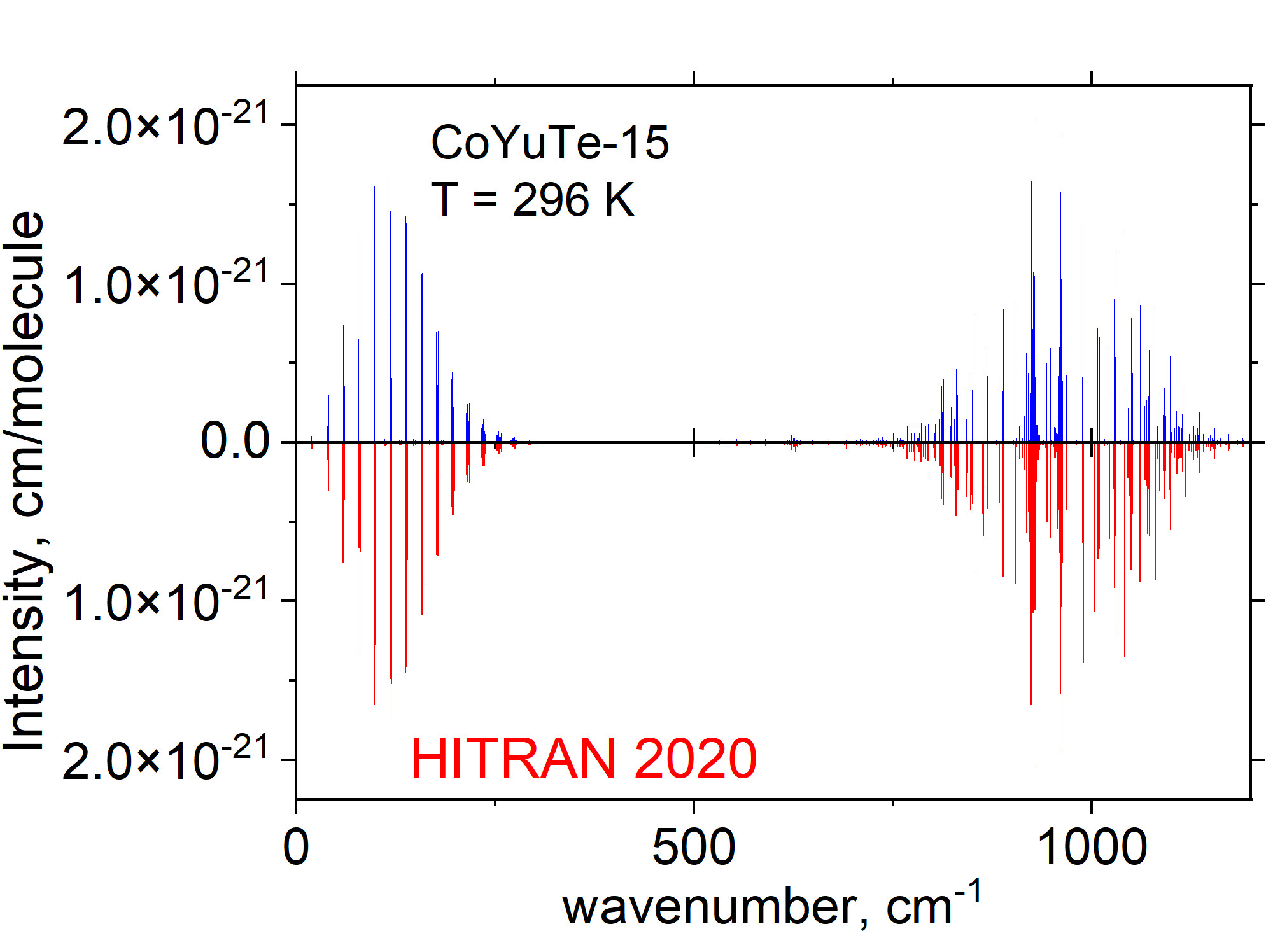}
\includegraphics[width=0.33\textwidth]{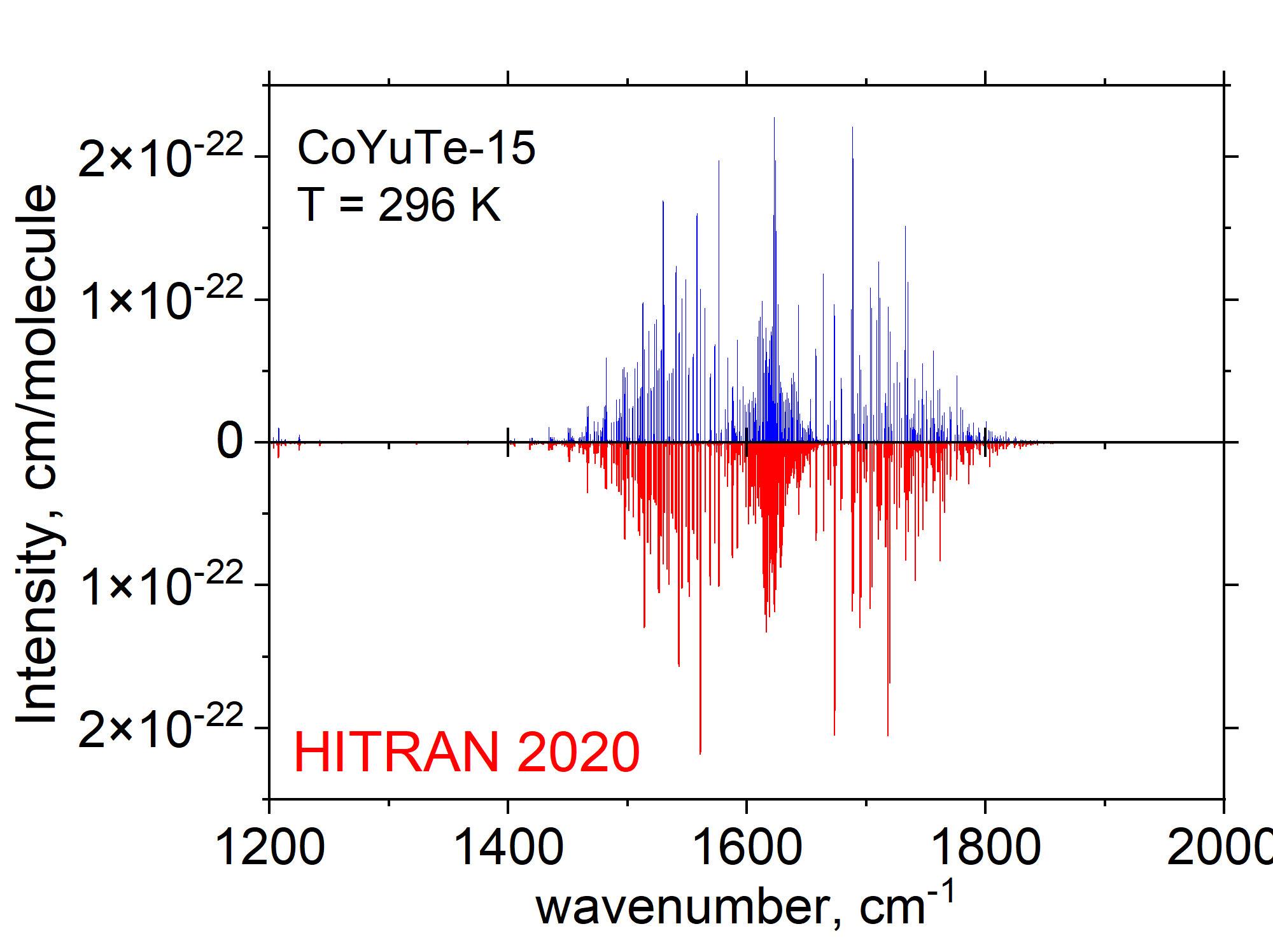}
\includegraphics[width=0.33\textwidth]{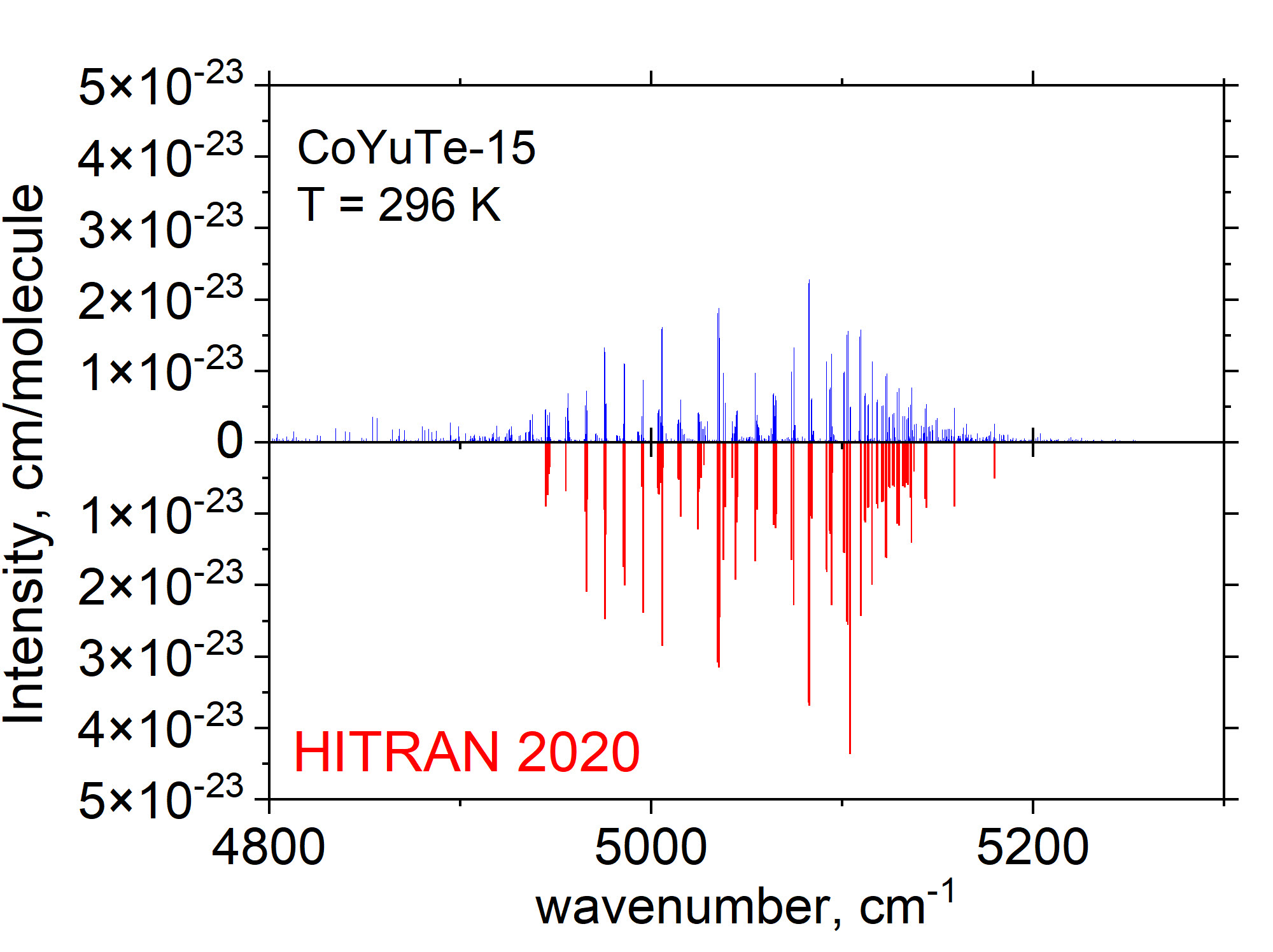}
\caption{\label{f:HITRAN:3:windows} Comparison of the \name\ spectrum with HITRAN~2000 \citep{HITRAN2020} at $T=296$~K. }
\end{figure}

In Figure~\ref{f:HITRAN}, an overview of the new line list is presented in a form of a room temperature ($296$~K) stick spectrum compared to the HITRAN~2020 \NH{3}{15} data \citep{HITRAN2020}. Using the MARVELised energies only, we obtain 53~856 \name\ transitions with intensities above $10^{-30}$~cm$^2$/molecule after scaling to terrestrial (natural) abundance as taken from \href{hitran.org}. This is to compare to {13~792} \NH{3}{15} transitions in HITRAN~2020, mostly covering the wavenumber region below \hl{3500}~\cm. Figure~\ref{f:HITRAN:3:windows} offers a more detailed comparison of the \name\ $T=296$~K spectrum with HITRAN~2020 in the three spectroscopic windows showing 4 bands, assuming natural abundances. This high resolution \NH{3}{15} \Marvel\ line list in the HITRAN format is provided as supplementary material.

\begin{table*}
\centering
\caption{\label{t:states} Extract from the \texttt{.states} file of the \NH{3}{15} \name\ line list. }
{\setlength{\tabcolsep}{1.5pt}
\scriptsize\tt
\begin{tabular}{rrrrrrcrrrrrcrrrrrrrrrrrrrrrrr}
\toprule \toprule
        $i$  &  \multicolumn{1}{c}{$\tilde{E}$/\cm}   &  $g$  &  $J$  & \multicolumn{1}{c}{unc/\cm} & \multicolumn{1}{c}{$\tau$/s} & $\Gamma_{\rm tot}$ & $n_1$ & $n_2$ & $n_3$ & $l_3$ & $n_4$ & $l_4$ & $\Gamma_{\rm vib}$ & $K$ & $\Gamma_{\rm rot}$   & $N$& s/a & $|C_i^{2}|$ &$v_1$ & $v_2$ & $v_3$ & $v_4$ & $v_5$ & $v_6$ & \multicolumn{1}{c}{Label}  &$\tilde{E}_{\rm calc.}$/\cm  \\
 \midrule
  30257 &    16.904164 &  12 &  1 &   0.000141 & 6.8781E+06 & E'  &  0 &  0 &  0 &  0 &  0 &  0 & A2"  &  1 &  E"  &  1   & a   &   1.00 & 0  &  0 &  0 &  0 &  0&  1 &  IE  &    16.905288\\
  30258 &   978.920185 &  12 &  1 &   0.000261 & 6.7761E-02 & E'  &  0 &  1 &  0 &  0 &  0 &  0 & A2"  &  1 &  E"  &  2   & a   &   1.00 & 0  &  0 &  0 &  0 &  0&  3 &  IE  &   978.913857\\
  30259 &  1637.547381 &  12 &  1 &   0.000461 & 2.2083E-01 & E'  &  0 &  0 &  0 &  0 &  1 &  1 &  E"  &  1 &  E"  &  4   & a   &   1.00 & 0  &  0 &  0 &  0 &  1&  1 &  IE  &  1637.563266\\
  30260 &  1643.458772 &  12 &  1 &   0.000424 & 2.1748E-01 & E'  &  0 &  0 &  0 &  0 &  1 &  1 &  E'  &  0 & A2'  &  4   & s   &   1.00 & 0  &  0 &  0 &  0 &  1&  0 &  IE  &  1643.466108\\
  30261 &  1886.697223 &  12 &  1 &   0.000424 & 3.9163E-02 & E'  &  0 &  2 &  0 &  0 &  0 &  0 & A2"  &  1 &  E"  &  3   & a   &   1.00 & 0  &  0 &  0 &  0 &  0&  5 &  IE  &  1886.683504\\
  30262 &  2553.937518 &  12 &  1 &   0.000616 & 5.2107E-02 & E'  &  0 &  1 &  0 &  0 &  1 &  1 &  E'  &  0 & A2'  &  6   & s   &   1.00 & 0  &  0 &  0 &  0 &  1&  2 &  IE  &  2553.902032\\
  30263 &  2590.898376 &  12 &  1 &   0.000848 & 5.1537E-02 & E'  &  0 &  1 &  0 &  0 &  1 &  1 &  E"  &  1 &  E"  &  6   & a   &   1.00 & 0  &  0 &  0 &  0 &  1&  3 &  IE  &  2590.901311\\
  30264 &  2891.605307 &  12 &  1 &   0.000616 & 2.3392E-02 & E'  &  0 &  3 &  0 &  0 &  0 &  0 & A2"  &  1 &  E"  &  5   & a   &   1.00 & 0  &  0 &  0 &  0 &  0&  7 &  IE  &  2891.589118\\
  30265 &  3199.900421 &  12 &  1 &   0.401000 & 4.0453E-02 & E'  &  0 &  2 &  0 &  0 &  1 &  1 &  E'  &  0 & A2'  &  7   & s   &   0.99 & 0  &  0 &  0 &  0 &  1&  4 &  Ca  &  3199.900421\\
  30266 &  3228.710631 &  12 &  1 &   0.201000 & 1.0889E-01 & E'  &  0 &  0 &  0 &  0 &  2 &  0 & A2"  &  1 &  E"  &  8   & a   &   0.99 & 0  &  0 &  0 &  0 &  2&  1 &  Ca  &  3228.710631\\
  30267 &  3254.708183 &  12 &  1 &   0.401000 & 1.0861E-01 & E'  &  0 &  0 &  0 &  0 &  2 &  2 &  E'  &  0 & A2'  &  9   & s   &   1.00 & 0  &  0 &  0 &  0 &  2&  0 &  Ca  &  3254.708183\\
  30268 &  3257.325527 &  12 &  1 &   0.401000 & 1.0759E-01 & E'  &  0 &  0 &  0 &  0 &  2 &  2 &  E"  &  1 &  E"  &  9   & a   &   0.99 & 0  &  0 &  0 &  0 &  2&  1 &  Ca  &  3257.325527\\
  30269 &  3350.280661 &  12 &  1 &   0.002106 & 1.0758E-01 & E'  &  1 &  0 &  0 &  0 &  0 &  0 & A2"  &  1 &  E"  &  10  & a   &   1.00 & 1  &  0 &  0 &  0 &  0&  1 &  IE  &  3350.267395\\
  \bottomrule\bottomrule
\end{tabular}}
\mbox{}\\
{\flushleft
\scriptsize
\begin{tabular}{ll}
\toprule\toprule
\noindent
$i$:& state identifier; \\
$\tilde{E}$:& state term value; \\
$g$:& state degeneracy; \\
$J$:& state rotational quantum number; \\
unc.:& energy uncertainty; \\
$\tau$:& lifetime;\\
$\Gamma_{\rm tot}$:& total symmetry in \Dh{3}(M); \\
$n_1$, $n_2$, $n_3$, $l_3$ $n_4$, $l_4$:& normal mode vibrational quantum numbers; \\
$\Gamma_{\rm vib}$:& symmetry of vibrational contribution in \Dh{3}(M); \\
$K$ :& rotational quantum number; \\
$\Gamma_{\rm rot}$:& symmetry of rotational contribution in \Dh{3}(M); \\
$N$: & vibrational state ID; \\
s/a: & inversion symmetry; \\
$|C_i^{2}|$:& largest coefficient used in the assignment; \\
$v_1 - v_6$:& \trove\ vibrational quantum numbers; \\
Label:& Label indicating if the term value is based on the \Marvel\ (`Ma')  \\
 & isotope extrapolation (`IE')  or the \name\ energy list (`Ca'); \\
$\tilde{E}_{\rm calc.}$:& original \name\ state term value. \\
\bottomrule
\end{tabular}
}
\end{table*}

\begin{table}
\centering
\caption{\label{t:trans}
Extract from a \texttt{.trans} file of the \NH{3}{15} \name\ line list. }
\tt
\centering
\begin{tabular}{rrrr}
\toprule\toprule
\multicolumn{1}{c}{$f$}	&	\multicolumn{1}{c}{$i$}	& \multicolumn{1}{c}{$A_{fi}$} \\
\midrule
1834174	&	2006561	&	4.8904E-05	\\
6245112	&	6543052	&	6.5693E-05	\\
2454658	&	3091139	&	3.7242E-12	\\
345773	&	154525	&	8.1452E+00	\\
3554160	&	3317840	&	9.4551E-19	\\
939596	&	816387	&	7.1870E-02	\\
8989567	&	9887870	&	1.0556E-17	\\
509601	&	419435	&	4.0892E-12	\\
\bottomrule\bottomrule
\end{tabular} \\ \vspace{2mm}
\rm
\noindent
$f$: Upper  state counting number;\\
$i$:  Lower  state counting number; \\
$A_{fi}$:  Einstein-$A$ coefficient (in s$^{-1}$).
\end{table}

\begin{figure}
\centering
\includegraphics[width=0.88\textwidth]{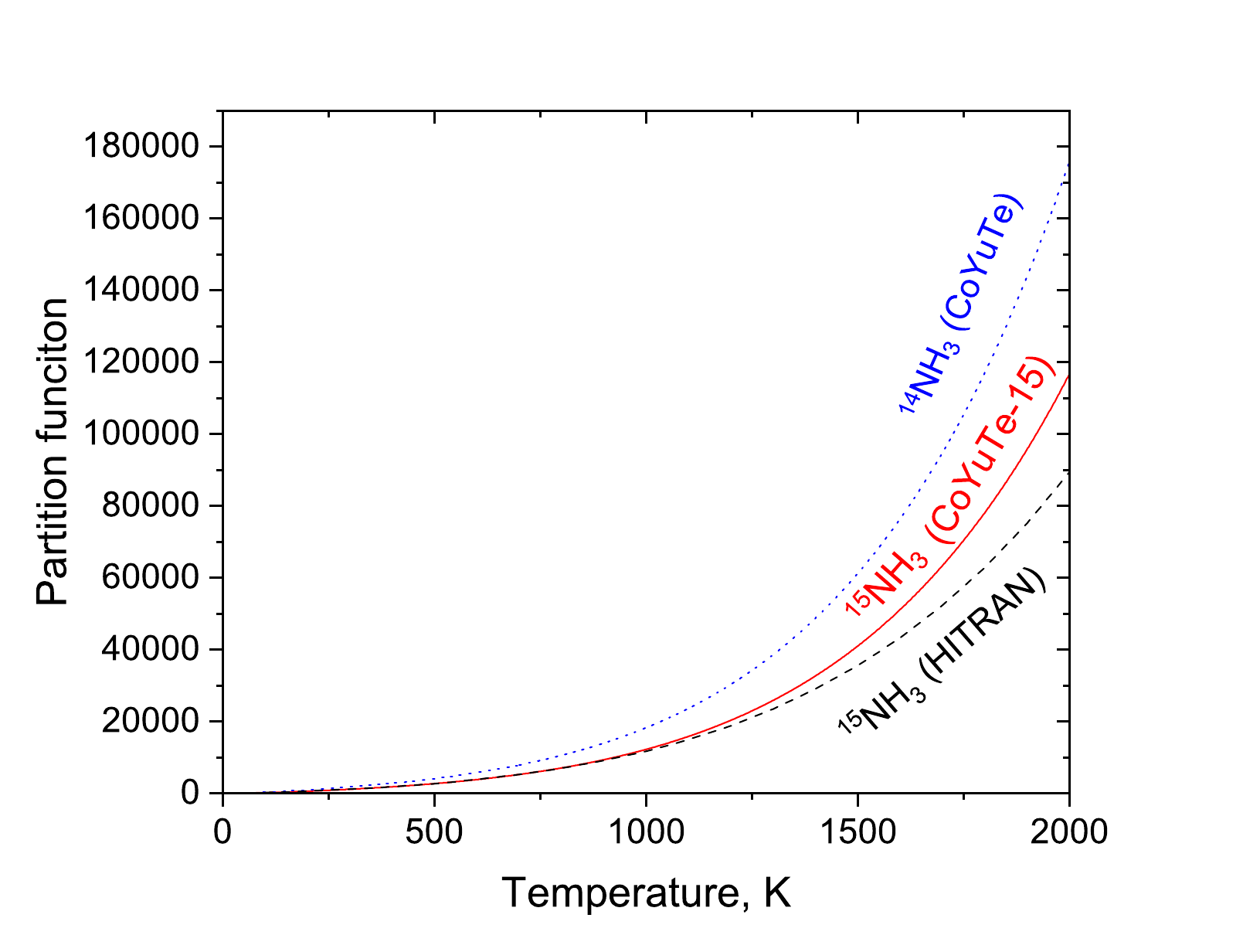}
\caption{\label{f:Q(t)} The partition function of \NH{3}{15} (\name) from this work (solid line) is compared to the HITRAN (TIPS) partition function of \NH{3}{15} \citep{TIPS2017} (dashed line) as well to the partition function of \NH{3}{14} from the CoYuTe line lists (dotted line).}
\end{figure}

Partition functions of \NH{3}{15} (CoYuTe-15) and \NH{3}{14} (CoYuTe) are compared in Fig.~\ref{f:Q(t)}, where we also show the HITRAN partition function of \NH{3}{15} from  TIPS  \citep{21GaViRe}. The latter is significantly smaller due to insufficient number of vibrational states used.

\begin{figure}
\centering
\includegraphics[width=0.88\textwidth]{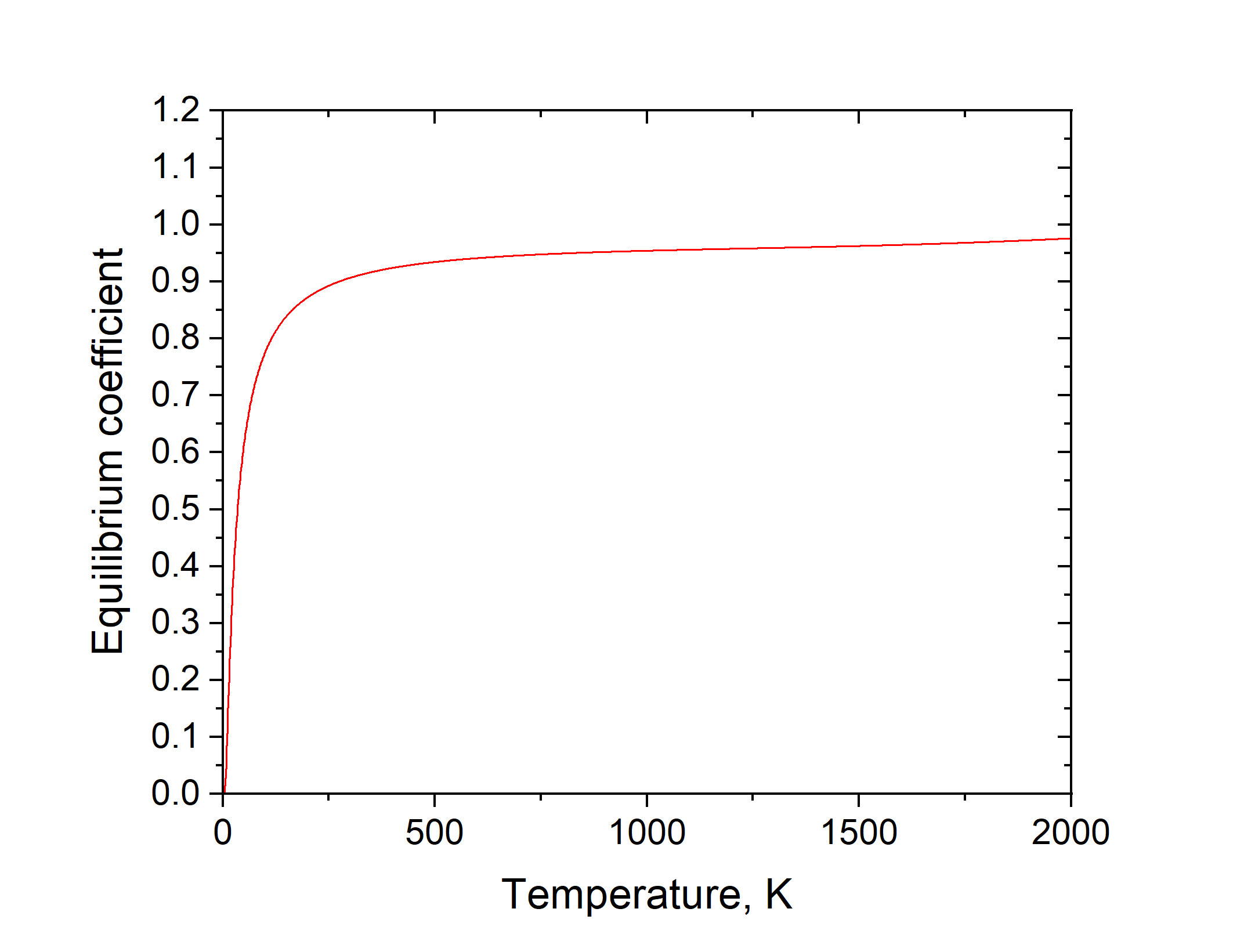}
\caption{\label{f:K(t)} Equilibrium constants of the reaction in Eq.~\eqref{e:itopization} as function of $T$.}
\end{figure}

Using these these two CoYuTe partition functions of \NH{3}{15} and \NH{3}{14}, we evaluated their equilibrium constants corresponding to the atomisation reaction
\begin{equation}
\label{e:itopization}
    ^{15}{\rm N}  + {}^{14}\text{NH}_3  \longrightarrow {}^{14}{\rm N}  + {}^{15}{\rm NH}_3.
\end{equation}
For the reaction
\begin{equation}
{\rm A} + {\rm B} \longrightarrow {\rm C} + {\rm D}
\end{equation}
we evaluate the temperature dependent equilibrium constant as follows \citep{jt344}
\begin{equation}
K = \left(
\frac{ m_{\rm A} m_{\rm B}}{m_{\rm C} m_{\rm D}} \right)^{3/2} \frac{Q_{\rm A} Q_{\rm B}}{Q_{\rm C} Q_{\rm D}} \exp\left(-\frac{c_2\tilde U}{T}\right),
\end{equation}
where $Q_{\rm A}$ and $Q_{\rm C}$ are atomic partition functions assumed to be 2 and 3 for $^{15}$N and $^{14}$N, respectively, $Q_{B}$ and $Q_{D}$ are the internal partition functions of \NH{3}{14} and \NH{3}{15}, respectively and $c_2$ is the second radiative constant (K/\cm).  The enthalpy of the reaction $\tilde{U}$  (\cm) is given by
$$
\tilde{U} = E_{0}^{\rm D} -  E_{0}^{\rm B},
$$
where $E_{0}^{\rm B}$ ( \NH{3}{14}) and  $E_{0}^{\rm D}$(\NH{3}{15}) are the molecular zero-point-energies,  8017.62~\cm\ and 8001.42~\cm, respectively. The temperature dependence of $K$ for this reaction is illustrated in Fig.~\ref{f:K(t)}. It is close to 1 at moderate and high temperatures but  exhibits drastic variation at very low values of $T$.

\begin{figure}
\centering
\includegraphics[width=0.7\textwidth]{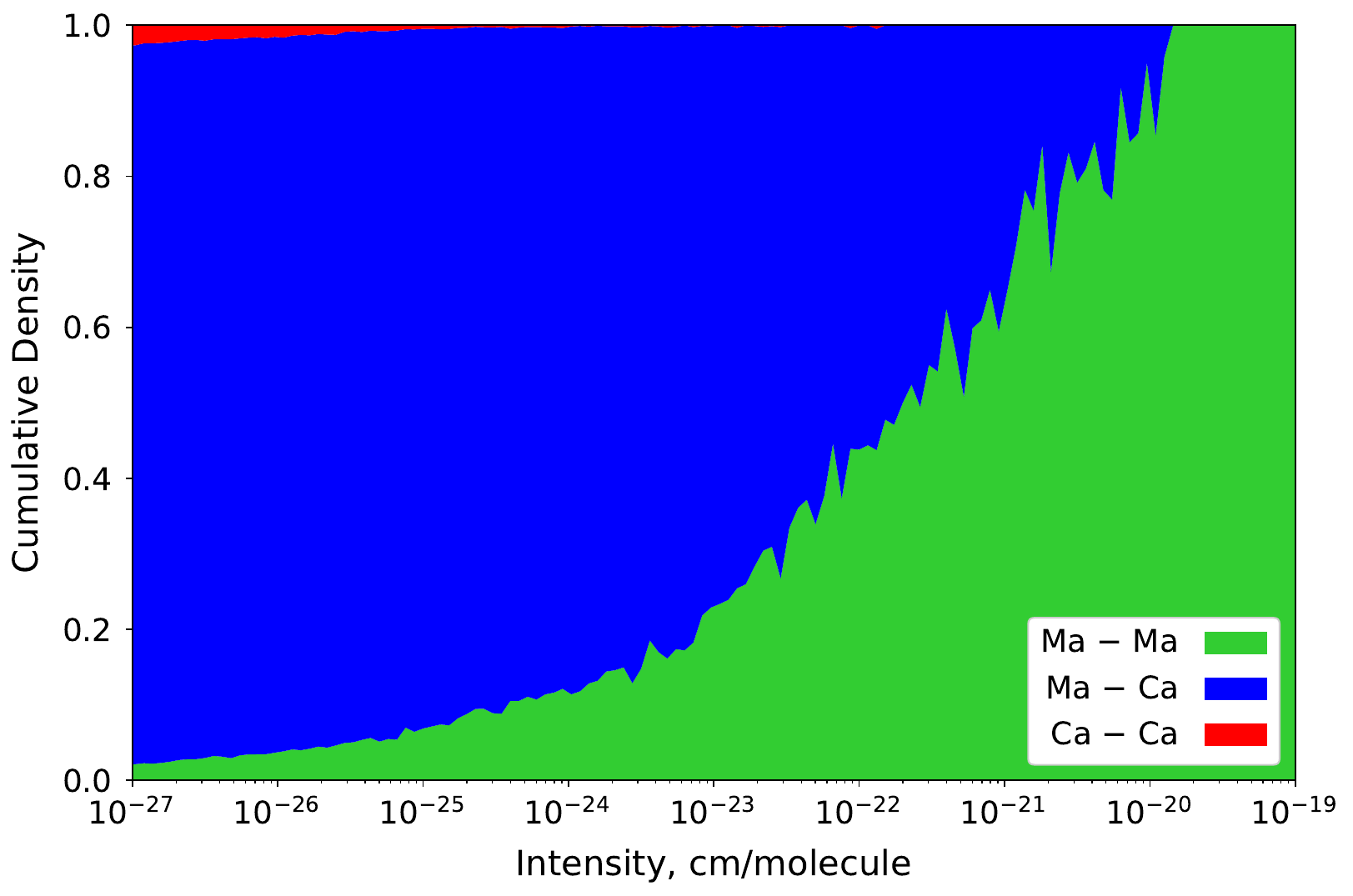}
\caption{\label{f:cint} The cumulative density of energy sources for each transition as a function of transition intensity at 296~K. The blue region groups together both Ma~$-$~Ca and Ca~$-$~Ma transitions. The \NH{3}{15} abundance of 1 was assumed.
}
\end{figure}

\begin{figure}
\centering
\includegraphics[width=0.7\textwidth]{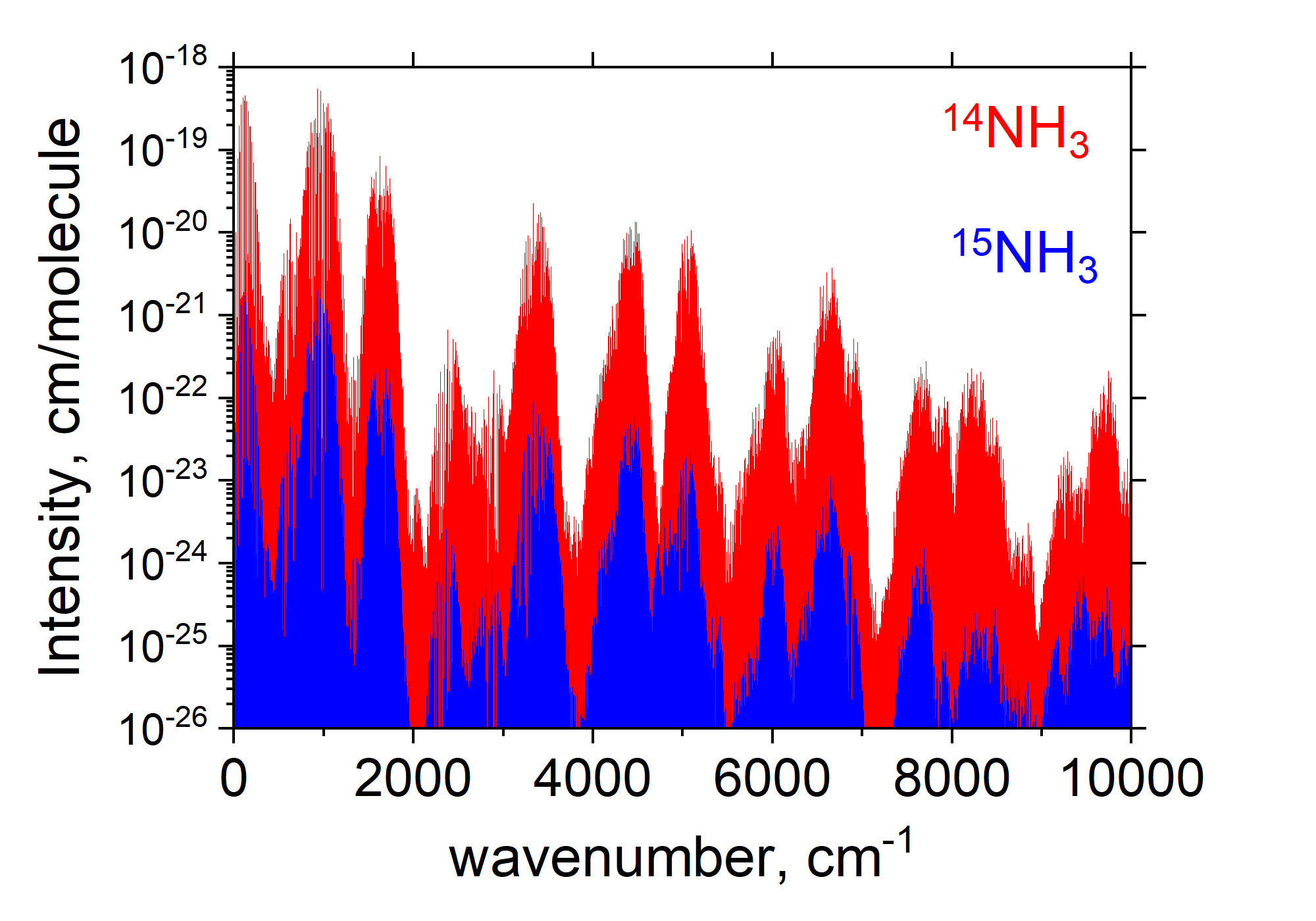}
\caption{\label{f:NH3-15vs14}  Room temperature intensities of \NH{3}{15} relative to that of \NH{3}{14}, scaled to the corresponding natural abundances, 3.66129$\times 10^{-3}$ and 0.995872, respectively. The \NH{3}{14} intensities are computed using the CoYuTe line list \citep{19CoYuTe}.}
\end{figure}

Figure~\ref{f:NH3-15vs14} shows how the intensities of \NH{3}{15} compare to \NH{3}{14} assuming natural abundances. The majority of the region is well above $10^{-26}$~cm/molecule.
In Figure~\ref{f:cint}, the cumulative density of energy sources for each transition is plot as a function of transition intensity for the \name\ line list at 296~K. This figure demonstrates that essentially all transitions with intensity above 10$^{-20}$ cm/molecule have both lower and upper energy levels accounted for in the \Marvel\ analysis. Even at 10$^{-22}$ cm/molecule, approximately half of the transitions are accounted for in the \Marvel\ analysis.

\begin{figure}
\centering
\includegraphics[width=0.7\textwidth]{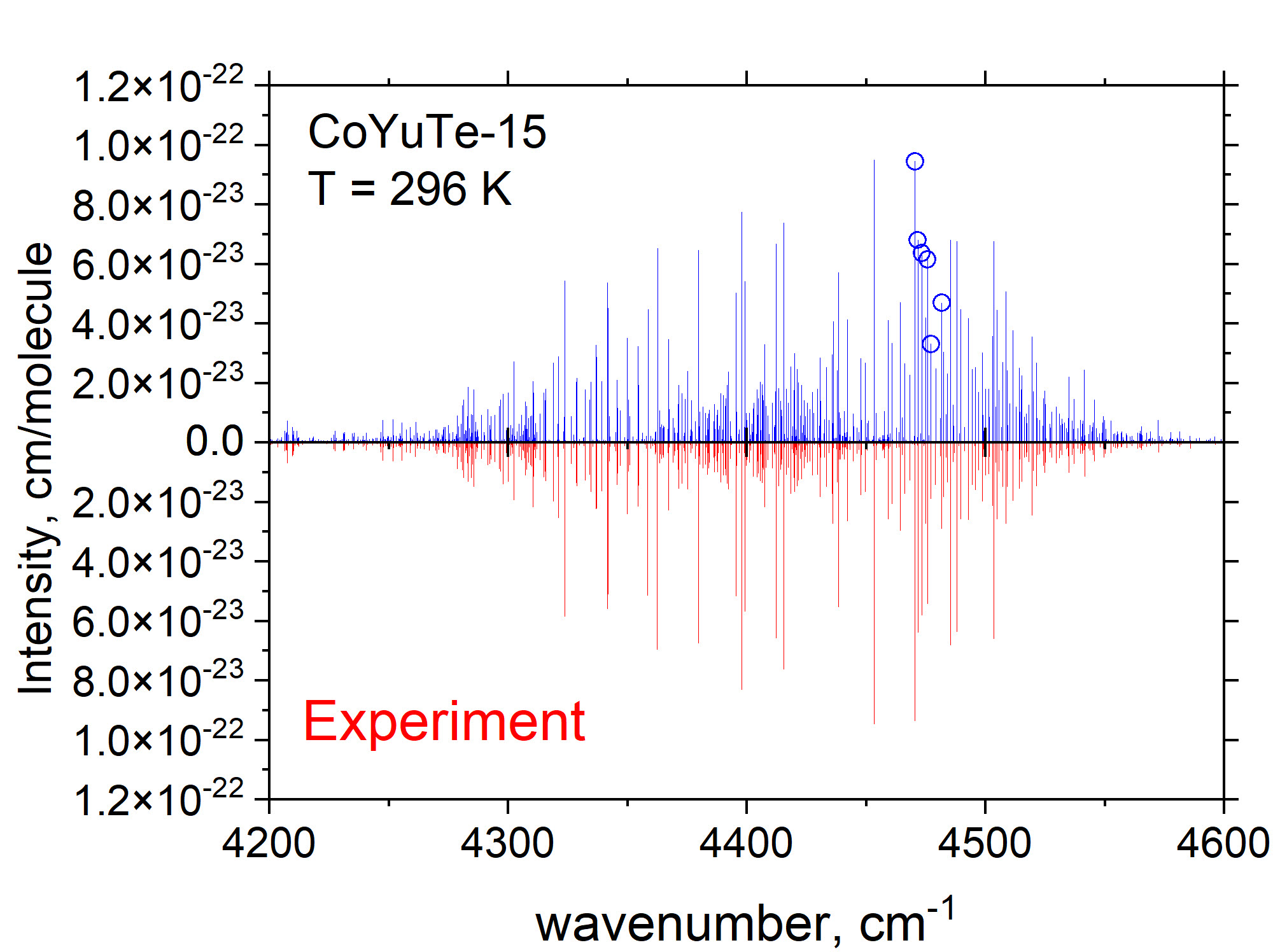}
\caption{\label{f:22CaCeVa}  Comparison of the \name\ spectrum in the 4400~\cm\ window with the unassigned experimental spectrum by \citet{22CaCeVa.NH3} at $T=296$~K. The theoretical spectrum is scaled to the  abundance by 5.7$\times 10^{-3}$ suggested by \citet{22CaCeVa.NH3}. Circles indicate lines selected for a detailed comparison in Table~\ref{t:obs/calc}  }
\end{figure}

\citet{22CaCeVa.NH3} reported spectra of ammonia in the 3900 and 4700 \cm\  region with a detailed analysis and assignment of \NH{3}{14} lines. The spectra contains also lines of \NH{3}{15}, which have not been assigned yet. In Fig.~\ref{f:22CaCeVa}, we compare our theoretical lines of \NH{3}{15} at $T=296$~K to the experimental spectral lines from \citet{22CaCeVa.NH3}. Here, the intensities were scaled to the abundance of $5.7\times 10^{-3}$ suggested by  \citet{22CaCeVa.NH3}. For a detailed analysis, six strong lines from the latter work were selected, analysed and compared to the \name\ values in Table~\ref{t:obs/calc}.  The agreement of the line positions is well within $0.001$~\cm\ for the MARVELsed values and within 0.02~\cm\ for the calculated ones. The intensity ratios vary from $\approx $75\% (weaker band $\nu_1+\nu_2$) to up to $\approx 15$\% (stronger band $\nu_2+\nu_3$). This agreement between theory and experiment shows that \name\ can be very useful as a starting point for formally assigning this spectrum.

\begin{table}
\centering
\caption{\label{t:obs/calc}  Comparison of line positions and intensities ($T=296$~K) for six selected lines from the experimental spectrum by \citet{22CaCeVa.NH3}, see lines indicated with circles in Fig.~\ref{f:22CaCeVa}. The theoretical spectrum is scaled to the  abundance by 5.7$\times 10^{-3}$ suggested by \citet{22CaCeVa.NH3}. The labels `Ma' and `Ca' indicate if the line position values are from MARVEL or calculated.
}
\centering
\begin{tabular}{llrrrccrrc}
\toprule\toprule
  &       & \multicolumn{4}{c}{Positions, \cm}   & &\multicolumn{3}{c}{Intensities, $10^{-23} $ cm/molecule} \\
  \cline{3-6}\cline{8-10}
Line   &   Band      & obs. & calc. & obs.-calc. & Ma/Ca && obs. & calc. & Ratio (calc./obs.) \\
\midrule
$^{R}R(3,3)  $& $\nu_2+\nu_3$    &    4470.52292 &     4470.52222 &   0.00070   &  Ma   &&  9.36 &   9.44 &  1.01\\
$^{R}R(4,3)  $& $\nu_2+\nu_3$    &    4471.65737 &     4471.67685 &  -0.01948   &  Ca   &&  6.39 &   6.80 &  1.06\\
$^{R}R(2,0)  $& $\nu_2+\nu_3$    &    4473.32228 &     4473.32191 &   0.00037   &  Ma   &&  5.82 &   6.37 &  1.10\\
$^{R}R(3,0)  $& $\nu_2+\nu_3$    &    4475.73941 &     4475.73901 &   0.00040   &  Ma   &&  5.43 &   6.16 &  1.13\\
$^{R}R(3,2)  $& $\nu_1+\nu_2$    &    4477.11327 &     4477.11322 &   0.00005   &  Ma   &&  1.89 &   3.31 &  1.75\\
$^{R}R(4,4)  $& $\nu_1+\nu_2$    &    4481.68651 &     4481.69588 &  -0.00937   &  Ca   &&  2.91 &   4.69 &  1.61\\
\bottomrule\bottomrule
\end{tabular}
\end{table}

\subsection{Collisional line-broadening}

Line-broadening parameters are an essential part of a spectroscopic database required to produce molecular cross sections. As part of the new line list for \NH{3}{15}, here we update the ammonia line line-broadening parameters applicable both to \NH{3}{14} and \NH{3}{15}. This is in line with the ExoMol line broadening `diet' adopted in \citet{jt684}. Collisional line-broadening of \NH{3}{14} has been extensively studied in the past, with a special attention to the microwave inversion spectra that exhibits its own peculiarities \citep{66Ben-Reuven.broad,22SkHaGo.broad}. A review of available broadening data in the infrared by molecular hydrogen and helium was made by HITRAN \citep{16WiGoKoHi.broad}.  Little vibrational dependence is observed for these perturbers, so only rotational dependence of broadening parameters is considered here. For H$_2$ broadening coefficients $\gamma$, a polynomial in $m$ and $K$, first appearing in \citep{04NeSuVa.NH3}, is adopted to describe the rotational dependence in terms of quantum number $K$ and rotational line index $m$. We introduce one change to this polynomial, namely using the absolute value of $\mid\!\! m\!\! \mid$ instead of $m$ to ensure correct behaviour in P- and R-branches. The spectroscopic coefficient $m$ is defined as $-J''$ for P-branch and $J''+1$ for  R-branch.

The single-power law is assumed for the temperature dependence of $\gamma(T)$:
\begin{equation}
\gamma(T)=\gamma_0(T_{\rm ref}) \left(\frac{T_{\rm ref}}{T}\right)^n,
\end{equation}
where $n$ is the temperature exponent and $\gamma_0(T_{\rm ref})$ is the half-width-at-half-maximum (HWHM) of the line at the reference temperature.

\begin{figure}
\centering
\includegraphics[width=0.7\textwidth]{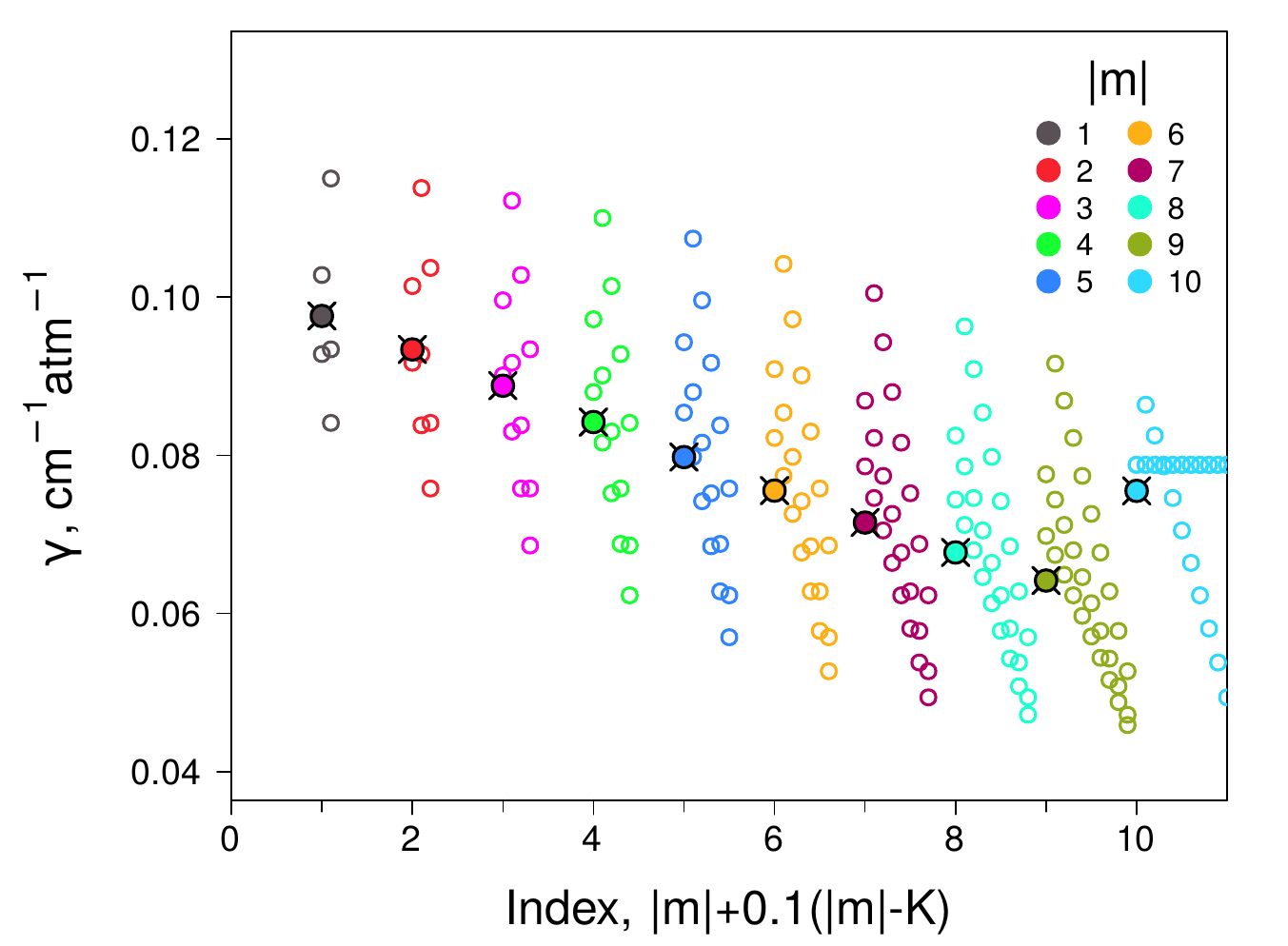}
\caption{
Collisional broadening parameters for the \NH{3}{}--H$_2$ system. Empty circles are HITRAN data for lines with various rotational assignments \{$m,K''$\}, filled-in circles are their $K$-averages (\texttt{m1} diet)
}
\label{f:m1diet}
\end{figure}

Here we provide two ExoMol broadening `diets' for \NH{3}{} in the form  of \textsf{.broad} files.  The first one uses the \texttt{m0} diet \citep{jt939} represented by the $m$-dependence only with dependencies on all other quantum numbers averaged. We also give a version with the \texttt{m1} diet, with $\gamma=\gamma(\mid \!\!m\!\!\mid,K'')$, i.e. $m$ and $K''$ dependent, which should provide a more accurate description of \NH{3}{} line shapes broadened by H$_2$.  Although broadening coefficients for various $K$ can substantially differ (see Fig.~\ref{f:m1diet}), the non-rigorous quantum rotational number $K$ is not always well characterised and therefore the \texttt{m0} diet can be more practical. Wherever this assignment is known and reliable, the \texttt{m1} diet can be used as a more complete representation of the broadening parameters.

Helium-broadening parameters show little variation even against  the rotational variable $m$, so the \texttt{m0} diet should suffice. The \texttt{m1} diet for this broadener is still provided for the sake of completeness.
To construct the \texttt{m}-diets helium-broadening, the HITRAN data \citep{16WiGoKoHi.broad} for $\gamma$ and $n$ were used. These data are based on \NH{3}{14} measurements, but should also work well for broadening of \NH{3}{15}. \citet{04NeSuVa.NH3} estimate that the broadening parameters for both isotopologues  are within 3\% at the measured temperatures; this is well below the uncertainties associated with the current parametrisations of $\gamma$ and $n$.

\section{Conclusions}

An extensive ExoMol line list \name\ for the isotopologue of \NH{3}{15} CoYuTe-15 is presented.  The line list was computed with the variational program \TROVE\  using the spectroscopic model (empirical PES and \ai\  DMS) of \NH{3}{} developed and used for production of a hot line list of \NH{3}{14} by \citet{19CoYuTe}. The line list covers the wavenumber range up to 10000~\cm\ and the rotational excitations up to $J=30$. In order to improve the variational energies, a set of \hl{2777}  empirically derived energy levels of \NH{3}{15} was constructed using the \Marvel\ procedure and used to replace the theoretical energies where possible. On top of that, we used an isotopologue extrapolation  technique  \citep{17PoKyLo} to generate energies for states not covered by the \NH{3}{15} \Marvel\ but featured in the \NH{3}{14} \Marvel\ set \citet{20FuCoTe}. This IE technique was applied to fill the gaps in the \Marvel\ energies from  the ground state, $\nu_2$, $2\nu_2$ and  $3\nu_2$. We show that the  line list is sufficiently accurate to help assign the existing experimental spectra of \NH{3}{15} as well to facilitate future spectroscopic analyses and astronomical detections. The line list as well as the spectroscopic model can be obtained at \url{www.exomol.com}.

\section*{Acknowledgements}
We thank Sammy Akkouche and Isabella Freitas for help at the start of the project.
This work was supported by the STFC Projects No. ST/R000476/1 and ST/Y001508/1. The authors acknowledge the use of the UCL Legion High Performance Computing Facility (Legion@UCL) and associated support services in the completion of this work, along with the Cambridge Service for Data Driven Discovery (CSD3) and  the DiRAC Data Intensive service DIaL2.5 at the University of Leicester, managed on behalf of the STFC DiRAC HPC Facility (www.dirac.ac.uk). These DiRAC services were funded by BEIS, UKRI and STFC capital funding and STFC operations grants. DiRAC is part of the UKRI Digital Research Infrastructure. This work was also supported by the European Research Council (ERC) under the European Union's Horizon 2020 research and innovation programme through Advance Grant number 883830.

\section*{Data Availability}

The data underlying this article are available in the article and in its online supplementary material. The line list and associated
data are available from \href{www.exomol.com}{www.exomol.com}. The codes used in this work, namely \textsc{TROVE} and \textsc{ExoCross}, are freely available via \href{https://github.com/exomol}{https://github.com/exomol}.

\section*{Supporting Information}

The following is provided as supporting information (1) the \Marvel\ input (transitions and segment) file, output (energy) file and an auxiliary `segment' file containing the units description of the transition frequencies from the original sources and (2) a list of HITRAN-formatted high resolution transitions of \NH{3}{15} computed at $T=296$~K using the \name\ intensities and the \Marvel\ energies.


\bsp	
\label{lastpage}

\end{document}